\documentclass[a4paper,onecolumn,aps,pra]{revtex4}
\usepackage[utf8]{inputenc}
\usepackage{amsmath} 
\usepackage{mathtools}
\usepackage{amsfonts} 
\usepackage{amssymb}
\usepackage{graphics}
\usepackage{graphicx}
\usepackage{float}
\usepackage{color}
\usepackage{braket}
\usepackage{tikz}
\usepackage{pgfplots}
\usepackage{fancyhdr}
\usepackage{bbold}
\usepackage{hyperref}
\usepackage{listings}             % Include the listings-package
\usepackage{psfrag}
\usepackage{leftidx}
\usepackage{colordvi}
\usepackage{bbm}
\usepackage[caption=false]{subfig}
\newcommand{\comment}[1]{}
\newcommand{\eq}[1]{\begin{equation}#1\end{equation}}
\newcommand{\dd}{\mathrm{d}}
\newcommand{\ee}{\mathrm{e}}

\newcommand{\ch}{\mathrm{ch}}
\newcommand{\sh}{\mathrm{sh}}

\newcommand{\const}{\mathrm{const}}
\newcommand{\ceff}{c_{\mathrm{eff}}}
\newcommand{\ceffp}{c'_{\mathrm{eff}}}
\newcommand{\twomat}[4]{\left(\begin{array}{cc} #1 & #2 \\ #3 & #4\end{array}\right)}

\newcommand{\lneg}{\mathcal{E}}
\newcommand{\lnf}{\mathcal{E}_f}
\newcommand{\lns}{\mathcal{E}_s}
\newcommand{\mi}{\frac{1}{2}\, \mathcal{I}_{1/2}}
\newcommand{\Tr}{\mathrm{Tr}}
\newcommand{\norm}[1]{\| #1 \|}
\newcommand{\asin}[1]{\arcsin{#1}}
\newcommand{\acos}[1]{\arccos{#1}}
\newcommand{\real}[1]{\mathrm{Re}(#1)}
\newcommand{\imag}[1]{\mathrm{Im}(#1)}

\pgfplotsset{compat=newest}

\usetikzlibrary{external}
\usetikzlibrary{calc,  shapes, backgrounds, arrows, chains, matrix, positioning, scopes,
                decorations, decorations.pathmorphing,patterns, fit, pgfplots.groupplots, plotmarks}
\tikzset{>=latex}
\usetikzlibrary{external}
\usetikzlibrary{external}
\begin{document}

\title{Time evolution of entanglement negativity across a defect}

\author{Matthias Gruber}
\affiliation{%
  Institute of Theoretical and Computational Physics,
  Graz University of Technology, NAWI Graz, Petersgasse 16, 8010 Graz, Austria
}%

\author{Viktor Eisler}
\affiliation{%
  Institute of Theoretical and Computational Physics,
  Graz University of Technology, NAWI Graz, Petersgasse 16, 8010 Graz, Austria
}%

\begin{abstract}
We consider a quench in a free-fermion chain by joining two
homogeneous half-chains via a defect. The time evolution
of the entanglement negativity is studied between adjacent segments
surrounding the defect. In case of equal initial fillings,
the negativity grows logarithmically in time and essentially equals
one-half of the R\'enyi mutual information with index $\alpha=1/2$ in the limit of large segments.
In sharp contrast, in the biased case one finds a linear increase followed by the
saturation at an extensive value for both quantities, which is due to
the backscattering from the defect and can be reproduced in a quasiparticle
picture. Furthermore, a closer inspection of the subleading corrections
reveals that the negativity and the mutual information have a small but
finite difference in the steady state.
Finally, we also study a similar quench in the XXZ spin chain via density-matrix
renormalization group methods and compare the results for the negativity
to the fermionic case.
\end{abstract}

%\tableofcontents
%\newpage

\maketitle

\section{Introduction}
The study of integrable quantum many-body systems out of equilibrium has received considerable
attention in the last decade \cite{calabrese2016introduction}.
Among the various research directions, the identification of the stationary ensemble \cite{VR16}
towards which the system locally relaxes after a quantum quench \cite{essler2016quench}
is of central importance. In understanding the mechanisms of this relaxation and equilibration,
a crucial role is played by entanglement dynamics. In particular, the buildup of an extensive
subsystem entropy can be traced back to the propagation of entangled quasiparticle excitations
created by the quench \cite{CC05,AC17}.

The situation is, however, less clear if the translational invariance of the Hamiltonian
governing the dynamics is broken by the presence of an impurity. In the generic case,
such a local deformation typically destroys the integrability of the model, which makes
the study of the dynamics far more complicated. Hence, most of our understanding
about entanglement evolution across a defect is essentially restricted to free-fermion systems.
These studies include free-fermion chains with a simple hopping defect
\cite{klich2009quantum,song2012bipartite,eisler2012onentanglement,GLC19},
the transverse Ising chain with a coupling defect \cite{igloi2009entanglement},
and even some more complicated interacting defects between non-interacting leads
\cite{KMV14,VS17,BM17,HM17}.

In general, the behaviour of the half-chain entanglement depends crucially on whether a
density bias is applied between the fermionic leads separated by the defect. For the unbiased case,
one finds a logarithmic growth of entanglement, with the same prefactor that governs
the equilibrium entropy scaling in the presence of a defect, which was studied both
numerically and analytically for hopping chains
\cite{peschel2005entanglementdefect,EP10,EG10,peschel2012exactresults,arias2019quantum}
as well as in the continuum \cite{calabrese2011entanglement,calabrese2012entanglement}.
The exact same relation can actually be found in conformal field theory (CFT) calculations
\cite{WWR18}, connecting entropy dynamics after the quench to the problem of ground-state
entanglement across a conformal interface \cite{sakai2008entanglement, brehm2015entanglement}.
In sharp contrast, entanglement was found to grow linearly if a density bias is applied to the
leads, which can be attributed to backscattering from the defect \cite{eisler2012onentanglement}.

The examples mentioned above deal with a bipartite setting, where the entanglement
entropy is calculated between two halves of the chain, with the defect located in the center.
It is natural to ask how the results are modified when entanglement is considered between
two finite segments surrounding the defect.
Since the reduced state of the two segments is not any more pure, the R\'enyi entropies
are no longer measures of entanglement. In such a tripartite geometry, a possible way
of quantifying entanglement is via the logarithmic negativity \cite{vidal2002computable,plenio2005logneg}.
In homogeneous systems, the time evolution of the entanglement negativity has already
been considered in a similar geometry for global \cite{coser2014entanglement,alba2019quantuminfo}
or local quenches \cite{wen2015entanglement,feldman2019dynamics},
and even for initial states with a temperature or density bias
\cite{eisler2014entanglementneg, hoogeveen2015entanglementneg,gullans2019entanglement}.
Particularly interesting are the recent results of Ref. \cite{alba2019quantuminfo},
where a nontrivial relation between the entanglement negativity and the
R\'enyi mutual information with index $\alpha=1/2$, measured between two neighbouring blocks,
has been established for free-particle systems after a global quench.

In the present paper we shall study the time evolution of the entanglement across a defect
by employing a fermionic version of the logarithmic negativity introduced in \cite{shapourian2017partial}.
One of the main questions we address is whether the relation between
entanglement negativity and R\'enyi mutual information still holds in such an inhomogeneous
quench problem. For the unbiased quench the answer is positive, with
the functional form of the entanglement evolution being well approximated
by that after a homogeneous local quench \cite{wen2015entanglement},
with an effective central charge that follows from the ground-state defect problem.
In case of a density bias, the leading order behaviour is again the same for both quantities
and can be captured by a semiclassical picture, similar to the one applied for the
half-chain entropy \cite{eisler2012onentanglement}.
Here the pairs of partially reflected and transmitted single-particle excitations
created at the defect contribute to a linear buildup of entanglement, which eventually
saturates at an extensive value. However, the subleading corrections for
the entanglement negativity and mutual information are found to differ
by a finite amount in the steady state, even in the limit of large segments.
Finally, we also study the standard logarithmic negativity for the XXZ chain with a
single coupling defect in a similar quench setup, using the numerical methods introduced in
\cite{ruggiero2016entanglement}. In particular, for the XX chain that is very closely
related to the hopping chain, we find that the spin-chain negativity is always
upper-bounded by the fermionic one.

The manuscript is structured as follows. In Section \ref{sec:methods} we introduce the
basic setup and the method of calculating the R\'enyi mutual information and
logarithmic negativity for fermionic Gaussian states. In Sec. \ref{sec:gs} we discuss
the entanglement negativity in the ground state of a chain with a defect.
Sections \ref{sec:quench1} and \ref{sec:quench2} are devoted to the study of the
quench from equal and unequal fillings, respectively. In Sec. \ref{sec:xxz} we compute
the negativity after a quench in the XXZ chain with a defect. We conclude with a
discussion of our results in \ref{sec:disc}, followed by two Appendices providing
some technical details of the calculations.

\section{Model and methods\label{sec:methods}}

We consider a chain of noninteracting fermions with a single hopping defect,
described by the Hamiltonian
\begin{equation}
 \hat H = \hat H_l + \hat H_r - \frac{\lambda}{2}\left( f_0^\dagger f_1 + f_1^\dagger f_0 \right) ,
  \label{eq: H}
\end{equation}
where the hopping amplitude $\lambda$ characterizes the defect in the middle of the chain,
while the homogeneous half-chains on the left/right hand side of the defect are given by
\begin{equation}
\hat H_l = -\frac{1}{2} \sum_{j = -N+1}^{-1} \left( f_j^\dagger f_{j+1} + f_{j+1}^\dagger f_j \right) \, ,
\qquad
\hat H_r = -\frac{1}{2} \sum_{j = 1}^{N-1} \left( f_j^\dagger f_{j+1} + f_{j+1}^\dagger f_j \right) \, .
  \label{eq: H_LR}
\end{equation}
The full chain has $2N$ sites and the fermionic annihilation (creation) operators
$f_j$ ($f_j^\dagger$) with $j=-N+1,\dots, N$ satisfy the canonical anticommutation relation
$\{ f_i, f_j^\dagger \} = \delta_{ij}$. The defect $\lambda \le 1$ is assumed to be weaker than
the tunneling in the leads $\hat H_l$ and $\hat H_r$.

\begin{figure}[H]
  \centering
  \includegraphics[width=0.55\textwidth]{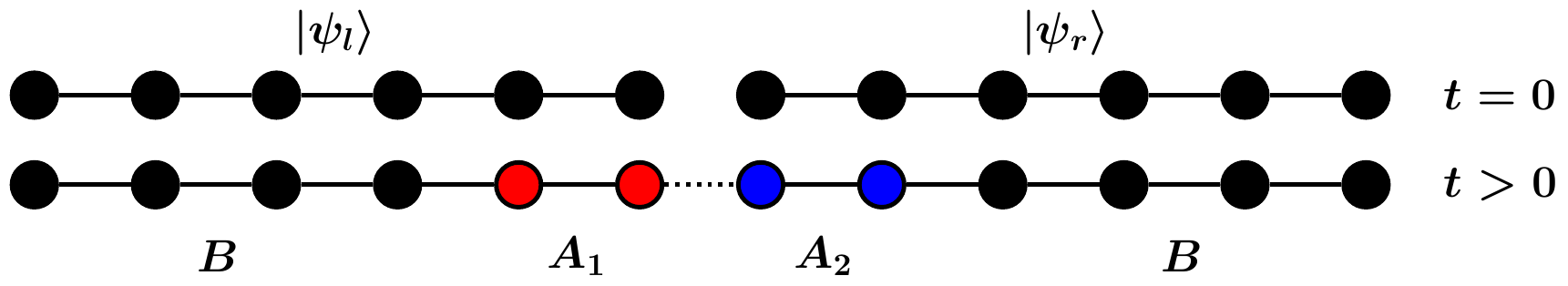}
  %\scalebox{1}{ \input{./PLOTS/TEX/setup.tex} } 
  \caption{Quench setup.}
  \label{fig: setup}
\end{figure}

In the following sections we shall either consider the ground state or the time evolution
generated by \eqref{eq: H}. In the latter case, the chain is initially
split in two halves and our quench protocol is depicted in Fig.~\ref{fig: setup}. 
Here the initial state $\ket{\psi_\sigma}$ of the left/right part ($\sigma=l, r$) is given
by the respective ground state of $\hat H_\sigma-\mu_\sigma \sum_{j\in\sigma}f_j^\dagger f_{j}$,
where the chemical potential $\mu_\sigma$ sets the filling $n_\sigma$ of the corresponding half-chain.
In particular, one sets $\mu_l = \mu_r = 0$ to initialize both chains in their half-filled ground states $n_l=n_r=1/2$,
whereas the choice $\mu_l = 1$ and $\mu_r = -1$ corresponds to the step-like density $n_l=1$ and $n_r=0$
(also known as the domain wall initial state).
In either case, the two halves are then coupled via a defect, depicted by the dashed bond in Fig.~\ref{fig: setup}, and
the resulting unitary time evolution is given by
\begin{equation}
  \ket{\psi(t)} = e^{-i \hat Ht} \ket{\psi_l} \otimes \ket{\psi_r} \, .
  \label{psit}
\end{equation}

We are primarily interested in the buildup of entanglement in the time-evolved state $\ket{\psi(t)}$,
between two segments $A_1$ and $A_2$ as shown by the colored sites in Fig.~\ref{fig: setup}.
We restrict ourselves to the case of adjacent segments of equal lengths $\ell$, located
symmetrically around the defect. The bipartite case of $\ell=N$ ($B=\emptyset$)
was studied in Ref. \cite{eisler2012onentanglement}, where the entanglement in the pure state
$\rho = \ket{\psi(t)} \bra{\psi(t)}$ is simply measured by the R\'enyi entropies between the two halves.
In general, however, one has to first extract the reduced density matrix
$\rho_{A} = \Tr_{B}(\rho)$ of the subsystem $A=A_1 \cup A_2$ by tracing out
over the environment $B$. This leaves us with a mixed state where the entanglement
is much harder to be quantified and requires a proper measure.

%%%%%%%%%%%%%%%%%%%%%%%%%%%%%%%%%%%%%%%%%
%%%%%%%%%%%%%%%%%%%%%%%%%%%%%%%%%%%%%%%%%

Before introducing this measure, one should remark that the time-evolved state \eqref{psit}
is Gaussian and thus fully characterized by its correlation matrix
$C_{mn}(t) = \bra{\psi(t)} f_m^\dagger f_n \ket{\psi(t)} $. Indeed, since we are dealing with
free fermions, the Hamiltonian \eqref{eq: H} can be written in the quadratic form
\eq{
\hat H = \sum_{m,n} H_{m,n}  f_m^\dagger f_{n} \, ,
}
which defines the elements $H_{m,n}$ of the hopping matrix $H$.
Then the time evolution of the correlation matrix can simply be obtained as
%, starting from the correlation matrix \cite{peschel2003calculation, eisler2018properties, eisler2012onentanglement} at $t = 0$, $C(0)$, 
%
\begin{equation}
  C(t) = e^{i Ht} \, C(0) \, e^{-i Ht} \, .
  \label{eq: C(t)}
\end{equation}
The initial correlation matrix of the decoupled system at $t=0$ has a block-diagonal form
\eq{C(0)=\twomat{C_l}{0}{0}{C_r}, \label{eq: blockform}}
where $C_\sigma$ is the ground-state correlation matrix at filling $n_\sigma$,
with matrix elements given by \cite{fagotti2011universal}
\begin{equation}
    (C_{\sigma})_{m,n} = \frac{1}{2(N+1)} \left[
    \frac{ \sin(q_{F,\sigma} (m-n)) }{ \sin(\frac{\pi}{2(N+1)} (m-n)) } -
    \frac{\sin(q_{F,\sigma} (m+n))}{ \sin(\frac{\pi}{2(N+1)} (m+n)) } \right],
\end{equation}
and the Fermi wavenumber is defined as
\eq{
q_{F,\sigma} = \frac{\pi(n_\sigma \, N +1/2)}{N+1} \, .
\label{qf}}
In particular, half-filling $n_\sigma = 1/2$ corresponds to a Fermi momentum $q_{F,\sigma} = \pi/2$,
whereas for the step initial condition ($n_l=1, \, n_r=0$) the correlations simplify to $(C_l)_{m,n}=\delta_{m,n}$
and $(C_r)_{m,n}=0$

\subsection{R\'enyi entropy and mutual information}

For the pure state $\rho = \ket{\psi(t)} \bra{\psi(t)}$, the reduced density matrix of
the segment $A_1$ is given by $\rho_{A_1} = \Tr_{B \cup A_2}(\rho)$.
The R\'enyi entropy between the segment $A_1$ and the rest of the system is defined as
\begin{equation}
   S_{\alpha}(\rho_{A_{1}}) = \frac{1}{1-\alpha} \ln \Tr \, (\rho^\alpha_{A_1}) \, ,
   \label{salpha}
\end{equation}
with the von Neumann entropy corresponding to the limit $\alpha=1$.
Note, however, that these measures do not give information about the entanglement
between $A_1$ and $A_2$. To gain some insight about the latter, one could consider 
the R\'enyi mutual information defined by the combination of subsystem entropies
\begin{equation}
  \mathcal{I}_\alpha%(\rho_{A})
   = S_\alpha(\rho_{A_1}) + S_\alpha(\rho_{A_2}) - S_\alpha(\rho_{A}) \, .
  \label{eq: mut_inf}
\end{equation}
The standard (von Neumann) mutual information with $\alpha=1$ is known
to be a measure of total (classical and quantum) correlations and is thus an upper bound
to the entanglement \cite{GPW05}. Unfortunately, however, for generic $\alpha$ it is not even a
proper measure. Indeed, it was demonstrated that $\mathcal{I}_\alpha$ may
become negative for $\alpha>2$ after a certain quench \cite{KZ17}.
On the other hand, it has been proved that $\mathcal{I}_\alpha$ is always positive
in the range $0<\alpha<2$ for both fermionic and bosonic Gaussian states \cite{CLE19}.
Furthermore, recently it was pointed out that the particular case $\alpha=1/2$ is
intimately related to a proper entanglement measure, the logarithmic negativity
(see below), after a global quench \cite{alba2019quantuminfo}.
Thus our focus will be exclusively on the case $\alpha=1/2$.

The R\'enyi mutual information \eqref{eq: mut_inf} is a simple combination of
bipartite entropies in the pure Gaussian state $\rho$ and is thus uniquely determined
by the correlation matrix \eqref{eq: C(t)}. In particular, for the segment $A_1$ one has
\begin{equation}
  S_\alpha(\rho_{A_1}) = 
  \frac{1}{1-\alpha} \sum_j \ln \left[ \zeta_j^\alpha + (1-\zeta_j)^\alpha \right] ,
  \label{eq: S_alpha}
\end{equation}
where $\zeta_j$ are the eigenvalues of the \emph{reduced} correlation matrix $C_{A_1}(t)$,
with indices restricted to the segment $m,n \in A_1$. Similar expressions hold for the other terms
in the mutual information \eqref{eq: mut_inf}, where the eigenvalues of the respective
reduced correlation matrices $C_{A_2}(t)$ and $C_{A}(t)$ must be used.

\subsection{Entanglement negativity}

The logarithmic negativity is a versatile measure of entanglement \cite{vidal2002computable}.
Its definition relies on the partial transpose of the density matrix, which may
have negative eigenvalues only if the system is entangled \cite{peres1996separability}.
For \emph{bosonic} systems, the effect of the partial transpose is well known to
be identical to a partial time reversal \cite{Simon00}. However, for \emph{fermionic} systems
this is not any more the case. Indeed, in contrast to partial time reversal \cite{shapourian2017partial},
the partial transpose in general does not lead to a Gaussian operator \cite{eisler2015onthept}.
However, it has been proved that the definition based on partial time reversal
also yields a proper measure of entanglement \cite{SR19}. Therefore, we shall adopt here
the fermionic version of the logarithmic negativity, since it is directly amenable
to correlation-matrix techniques.

In order to define the fermionic negativity, it is more convenient
to work in the Majorana operator basis
\begin{equation}
  c_{2j-1} = f_j + f_j^\dagger, \qquad c_{2j} = i (f_j - f_j^\dagger ) \, ,
  \label{majorana}
\end{equation}
satisfying the anticommutation relations $\{ c_m, c_n \} = 2\delta_{mn}$.
We can now expand the reduced density matrix $\rho_A$ supported on
the segment $A = A_1 \cup A_2$ encompassing the defect (see Fig.~\ref{fig: setup}) as
\begin{equation}
    \rho_A = \sum_{\substack{\kappa, \tau \\ |\kappa| + |\tau|= \text{even}}}
    w_{\kappa,\tau} c_{-2\ell+1}^{\kappa_{1}} \dots c_{0}^{\kappa_{2\ell}} 
  c_{1}^{\tau_{1}} \dots c_{2\ell}^{\tau_{2\ell}} \, .
\end{equation}
Here $\kappa_j$ and $\tau_j$ with $j=1,\dots,2\ell$ are bit strings associated to
the subspaces $A_1$ and $A_2$, with their norms defined as
$|\kappa| = \sum_j \kappa_j$ and $|\tau| = \sum_j \tau_j$, respectively.
The bit strings indicate whether a Majorana operator is included or not,
$c_j^0 = \mathbb{1}$ or $c_j^1 = c_j$, with the weight of the corresponding term in the expansion
given by $w_{\kappa,\tau}$. Importantly, the sum is restricted to terms, where the overall
number of Majorana operators is even, reflecting the global fermion-number parity symmetry of the state.

The partial time reversal $R_2$ with respect to $A_2$ acts as \cite{shapourian2017partial}
\begin{equation}
  \rho_A^{R_{2}} = O_+ =
  \sum_{\substack{\kappa, \tau \\ |\kappa| + |\tau|= \text{even}}} w_{\kappa,\tau}  i^{|\tau|}
   c_{-2\ell+1}^{\kappa_{1}} \dots c_{0}^{\kappa_{2\ell}} c_{1}^{\tau_{1}} \dots c_{2\ell}^{\tau_{2\ell}} \, ,
   \label{ptr}
\end{equation}
where we have introduced the shorthand notation $O_+$ which will be useful
also for the definition of the partial transpose. Note that, in general, $O_+$ is not
a Hermitian operator and its conjugate will be denoted by $O_-=O_+^\dag$.
The fermionic logarithmic negativity is then defined as
\begin{equation}
  \lnf = \ln \norm{\rho_A^{R_{2}}}_1=\ln \Tr \sqrt{O_+ O_-} \, .
  \label{lnf}
\end{equation}

Our goal is now to calculate $\lnf$ via the Majorana covariance matrix
\begin{equation}
  \Gamma_{mn} = \frac{1}{2} \Tr \left( \rho \left[ c_m, c_n \right] \right)  ,
\end{equation}
where $\rho$ is the density matrix obtained from the pure state \eqref{psit} as before.
Since the dynamics conserves the fermion number, the covariance matrix $\Gamma$
is completely determined by $C(t)$ obtained from \eqref{eq: C(t)}. Using the definition \eqref{majorana},
it is easy to show that the following relations hold
%
%\begin{equation}
    %i \Gamma_{2j-1,2l}= -i \Gamma_{2j,2l-1} = \delta_{jl} -2 \, \real{C_{jl}(t)} \, ,
    %\qquad
    %i \Gamma_{2j-1,2l-1} = i \Gamma_{2j,2l} = -2 \imag{C_{jl}(t)} \, .
    %\label{gammac}
%\end{equation}
\begin{equation}
    \Gamma_{2j-1,2l}= - \Gamma_{2j,2l-1} = i \left(2 \, \real{C_{jl}(t)} -\delta_{jl} \right) \, ,
    \qquad
    \Gamma_{2j-1,2l-1} = \Gamma_{2j,2l} = 2i \, \imag{C_{jl}(t)} \, .
    \label{gammac}
\end{equation}
Note that we have suppressed the explicit $t$-dependence of the $\Gamma$ matrix
for notational simplicity.
Due to the Gaussianity of the state, the reduced density matrix $\rho_A$ is
characterized by the reduced covariance matrix $\Gamma_A$. Moreover,
one can show that $O_\pm$ are both Gaussian operators, with corresponding
covariance matrix elements 
\eq{
(\Gamma_{\pm})_{mn} = \frac{1}{2} \Tr \left( O_{\pm} [c_m,c_n] \right) ,
}
that can be written in the block form \cite{eisler2015onthept,shapourian2019twisted}
\eq{\Gamma_\pm = 
\twomat{\Gamma_{A_1A_1}}{\pm i \, \Gamma_{A_1A_2}}
{\pm i \, \Gamma_{A_2A_1}}{-\Gamma_{A_2A_2}},
}
where the block indices $A_1$ and $A_2$ denote the corresponding submatrices of $\Gamma_A$.

Clearly, evaluating the entanglement negativity in \eqref{lnf} boils down to
an exercise of multiplying Gaussian operators and taking their trace.
This has been carried out in Ref. \cite{eisert2018entanglement} by
introducing the auxiliary density matrix
\begin{equation}
  \rho_\times = \frac{O_+ O_-}{\Tr \, (O_+ O_-)} ,
\end{equation}
which is a normalized Gaussian state with a real spectrum. Using the multiplication
rules of Gaussian states, one can show that the corresponding covariance matrix
can be written as \cite{eisert2018entanglement}
\begin{equation}
  \Gamma_\times \simeq \left(\frac{\mathbb{1}+\Gamma_A^{\, 2}}{2}\right)^{-1}
  \twomat{\Gamma_{A_1A_1}}{0}{0}{-\Gamma_{A_2A_2}},
  \label{gammax}
\end{equation}
where $\simeq$ denotes equality up to a similarity transformation. Indeed,
it turns out that the result for $\lnf$ depends only on the spectra $\{\pm \nu_j^\times\}$
of $\Gamma_\times$ as well as that $\{\pm \nu_j\}$ of $\Gamma_A$, where 
$j=1, \dots, 2\ell$ and the eigenvalues come in pairs due to the antisymmetry of the
covariance matrix. The traces appearing in $\lnf$ can then be evaluated as
\begin{equation}
  \Tr \, (O_+ O_-) = \Tr \, (\rho^2_A)
  = \prod_{j=1}^{2\ell} \frac{1 + \nu_j^2}{2} \, ,
  \qquad
  \Tr \, (\rho^{1/2}_\times) = 
  \prod_{j=1}^{2\ell} \left[ \left( \frac{1+\nu^{\times}_{j}}{2} \right)^{1/2} 
  + \left( \frac{1-\nu^{\times}_{j}}{2} \right)^{1/2} \right] .
  \label{trace}
\end{equation}
Finally, using the formula \eqref{salpha} for the R\'enyi entropies,
the logarithmic negativity can be put in the suggestive form
\eq{
\lnf = \frac{1}{2} \left[S_{1/2}(\rho_\times)-S_{2}(\rho_A)\right] .
\label{lnfs}}

We have thus obtained $\lnf$ as a combination of R\'enyi entropies of the
reduced density matrix $\rho_A$ and the auxiliary density matrix $\rho_\times$,
which in turn can be evaluated using the trace formulas \eqref{trace}.
Note that, since the covariance matrix $\Gamma_\times$ is equivalent
to the one in \eqref{gammax} which depends only on the matrix elements of
$\Gamma_A$, the negativity is uniquely determined by the fermionic correlation
matrix via \eqref{gammac}. It is instructive to check the limit when $\rho_A$
corresponds to a pure state, such that $\Gamma^2_A=\mathbb{1}$ and
one has trivially $S_2(\rho_A)=0$. Furthermore, from \eqref{gammax} one
observes that $\Gamma_\times$ becomes block diagonal and thus
$S_{1/2}(\rho_\times)=S_{1/2}(\rho_{A_1})+S_{1/2}(\rho_{A_2})$.
Substituting into \eqref{lnfs} and using the symmetry property
$S_{1/2}(\rho_{A_1})=S_{1/2}(\rho_{A_2})$ of the R\'enyi entropy,
one obtains the well known relation $\lnf = S_{1/2}(\rho_{A_1})$.

\section{Ground state entanglement\label{sec:gs}}

Although the main focus of our work is the time evolution of the entanglement
across a defect, it turns out to be very useful to have a look at the ground-state
entanglement first. Namely, we shall consider here the ground state of the chain
\eqref{eq: H} and calculate the entanglement negativity for the same geometry
as for the quench shown in Fig.~\ref{fig: setup}, i.e. for two equal segments
surrounding the defect. In fact, in the bipartite case when the segment is taken
to be the half-chain $(B=\emptyset)$, it has been shown that the entanglement
entropies in the ground state and after the quench are very closely related
\cite{eisler2012onentanglement,WWR18}.

The defect problem for the entanglement was first studied in Ref.
\cite{peschel2005entanglementdefect} where a single interval neighbouring the
defect was considered in an infinite chain. The logarithmic scaling of the entropy
was found to persist, albeit with a prefactor (dubbed as effective central charge)
that varies continuously with the defect strength. Importantly, the contributions
to the entanglement from the two boundaries of the interval were found to be additive.
An analytical expression for the defect contribution was later derived in \cite{EP10}
by considering the half-chain entropy for $\alpha=1$, and further exact results
for various other $\alpha$ were obtained in \cite{peschel2012exactresults}.

We shall now argue that the effective central charge for $\alpha=1/2$ will govern
also the scaling of the entanglement negativity. Obviously, for the bipartite case
this follows immediately from the relation $\lnf = S_{1/2}(\rho_{A_1})=\mi$.
However, even in the generic tripartite case, one expects that the negativity
should only be sensitive to the defect contribution. Indeed, $\lnf$ measures the
entanglement between the segments $A_1$ and $A_2$ and should not care
about the contribution of the homogeneous boundaries between $A$ and $B$.
Now, this is exactly the contribution contained in $S_{1/2}(\rho_{A})$, which
is subtracted in the mutual information. Indeed, as shown in
\cite{ossipov2014entanglement}, in the limit of $\ell \gg 1$, the R\'enyi
entropy of an interval containing the defect in the middle is just given by
the homogeneous result. Therefore, in complete analogy to
\cite{alba2019quantuminfo}, we assume by a continuity argument that the
relation $\lnf \simeq \mi$ should hold in the tripartite case as well. Note that the
factor $1/2$ just compensates the double counting of the defect contribution.

%%%%%%%%%%%%%%%%%%%%%%%%%%%%%%%%%%%%%%%%%
\begin{figure}[H]
    \hspace*{-0.2cm}
    %\scalebox{1}{\input{./PLOTS/TEX/1GS_ceff_kappa.tex}}\\
    \subfloat{\includegraphics{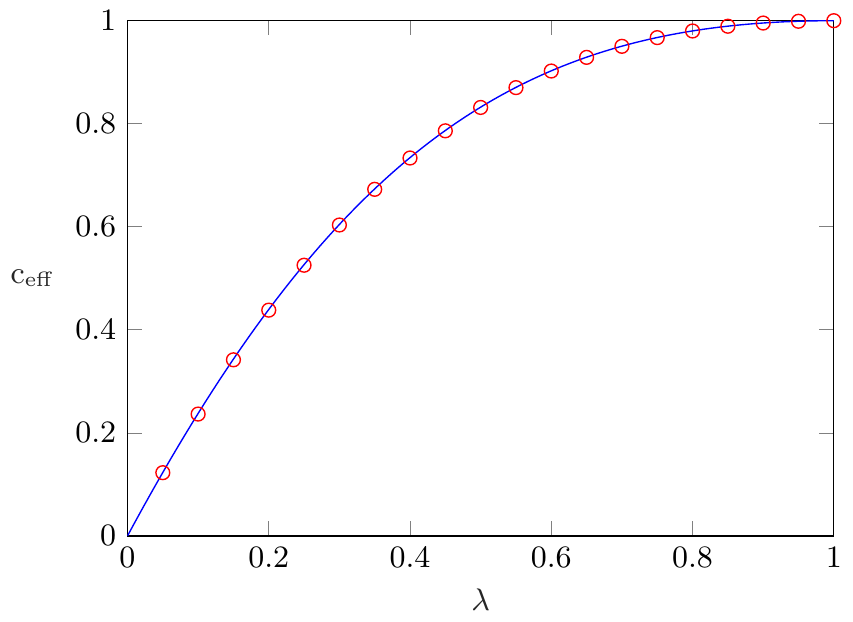}}
    \subfloat{\includegraphics{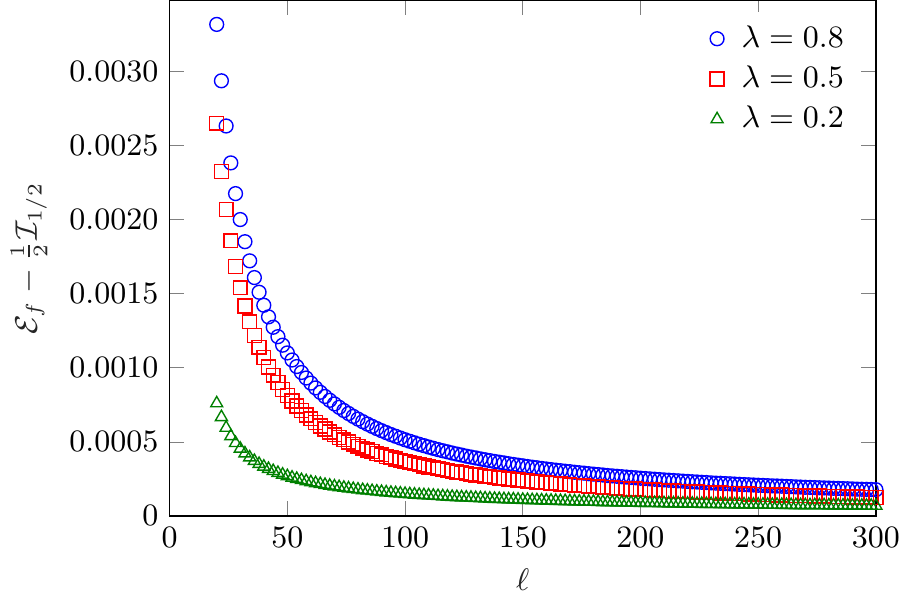}}
    \caption{Left: comparison of the fits for $\ceff$ (circles) according to Eq.~\eqref{eq: 1GS fit}, 
    to the analytical formula \eqref{ceff} (line). Right: difference of $\lnf$ and $\frac{1}{2}\mathcal{I}_{1/2}$ 
    as a function of $\ell$ for different coupling strengths $\lambda$.}
    \label{fig:gs}
\end{figure}
%%%%%%%%%%%%%%%%%%%%%%%%%%%%%%%%%%%%%%%%%

One can now use the results for the $\alpha=1/2$ R\'enyi entropy to put forward
the ansatz for the negativity
\eq{
\lnf \simeq \mi = \frac{\ceff}{4} \ln(\ell) + \const ,
\label{lngs}}
where the effective central charge reads \cite{peschel2012exactresults}
\begin{equation}
  \ceff = \frac{4}{\pi^2} \asin(s) (\pi - \asin(s)) \, ,
  \qquad
  s = \frac{2}{\lambda + \lambda^{-1}} \, .
  \label{ceff}
\end{equation}
Here $s$ is the transmission amplitude of the defect at the Fermi level $q_F=\pi/2$,
i.e. the square root of the transmission coefficient $s=\sqrt{T_{q_F}}$, see \eqref{Tq}.
Note that $\ceff$ is given by a smooth function that varies between zero and one,
which are the limiting cases of decoupled ($\lambda=0$) and homogeneous ($\lambda=1$) chains.
%In fact, the prefactor of the logarithm in \eqref{lngs} was chosen to reproduce the
%CFT result $c/4$ for the $c=1$.

To test our ansatz in \eqref{lngs}, we carried out numerical calculations using the methods
of Sec. \ref{sec:methods}, where $C(t)$ has to be replaced by the ground-state correlation matrix.
This can be evaluated directly in the thermodynamical limit $N \to \infty$ \cite{peschel2005entanglementdefect},
with the formulas summarized in Appendix \ref{app:corr}.
For a fixed value of $\lambda$, we find indeed a logarithmic growth of $\lnf$ with the segment size,
which can be fitted to 
\begin{equation}
  \lnf = \frac{\ceff}{4} \ln(\ell) + a + \frac{b}{\ell} \, ,
  \label{eq: 1GS fit}
\end{equation}
in the range up to $\ell=300$ and including also subleading corrections.
A comparison of the fits for $\ceff$ obtained from Eq.~\eqref{eq: 1GS fit}
and the analytic prediction \eqref{ceff} is shown on the left of Fig.~\ref{fig:gs},
with a perfect agreement. Furthermore, we also compared the logarithmic negativity
and the mutual information directly, with their difference shown on the right of
Fig.~\ref{fig:gs}. One can clearly see that the difference
decreases with increasing $\ell$, which confirms the assumption $\lnf\simeq\mi$
up to subleading corrections that seem to vanish in the $\ell \to \infty$ limit.

\section{Quench from equal fillings\label{sec:quench1}}

After having investigated the ground-state problem, we now move to the quench scenario
depicted in Fig.~\ref{fig: setup}. First we consider equal fillings, restricting ourselves
to the case of half-filled chains $n_l=n_r=1/2$.
For a homogeneous chain ($\lambda=1$), the time evolution of the entanglement entropy
after such a local quench has been calculated for hopping chains \cite{EP07} as well as
within CFT \cite{CC07,SD11}. Moreover, CFT calculations could even be extended to
the treatment of the negativity after the quench, using the techniques introduced in \cite{CCT12,CCT13}.
For symmetric intervals in an infinite chain $N\to\infty$ and $t < \ell$, one obtains the result
\cite{wen2015entanglement}
\begin{equation}
  \lnf = \frac{c}{4} \ln \left( \frac{t^2+\epsilon^2}{\epsilon^2} \frac{\ell-t}{\ell+t} \right) + \text{const},
  \label{lnlqh}
\end{equation}
where $\epsilon$ is a short-distance cutoff, ubiquitous in CFT calculations.
Remarkably, the exact same result can be found for the R\'enyi mutual information $\mi$
in the limit of adjacent intervals, based on the the results of Ref. \cite{AB14},
where only the case $\alpha=1$ was considered but the generalization to $\alpha=1/2$
is trivial. It should be stressed that both CFT results contain, in general, a contribution from
a non-universal function which, however, is expected to vanish for the Dirac fermion theory
we are interested in, and is thus not included in \eqref{lnlqh}. The characteristic feature of
\eqref{lnlqh} is an early logarithmic growth for $t \ll \ell$ which then levels off into a plateau,
followed by a sharp decrease around $t \to \ell$. For $t>\ell$, i.e. after the front created
by the quench travels through the segment, the negativity assumes its ground-state value
$\lnf = c/4 \ln \ell + \const$.

On the other hand, for local quenches across a defect, another interesting result was found
for the time evolution of the R\'enyi entropy of a half-chain, $\ell=N$.
Namely, for a hopping chain with a defect, it turns out that the growth of the entropy is
logarithmic and governed by the exact same effective central charge as found for the
ground-state entanglement \cite{eisler2012onentanglement,TF15}. The result was later generalized
to arbitrary CFTs with a conformal defect \cite{WWR18}.
In particular, for $\alpha=1/2$ and in the limit $t \ll N$, one finds for the hopping chain
\eq{
S_{1/2}(\rho_{A_1}) = \frac{\ceff}{2}\ln (t) + \const,
\label{slq}}
where $\ceff$ is given by \eqref{ceff}. Note that in the homogeneous case $\ceff=1$,
and thus taking the limit $\ell \to \infty$, $t \gg \epsilon$ and setting $c=1$ in \eqref{lnlqh}
exactly reproduces \eqref{slq}, as it should.

We now consider the negativity in the tripartite setup of Fig.~\ref{fig: setup}.
Similarly to the ground-state case in Sec. \ref{sec:gs}, we argue that the
relation $\lnf \simeq \mi$ should hold also for the local quench across a defect.
However, apart from the homogeneous case $\lambda=1$, we are not aware of
any calculations (neither lattice, nor CFT), which would generalize the formula
\eqref{slq} on the R\'enyi entropy for an interval $\ell<N$ that is not the half-chain.
Nevertheless, it is reasonable to expect that, until the front reaches the boudary $t < \ell$,
the entropy of the composite interval $S_{1/2}(\rho_A)$ actually remains constant.
Then, by combining the results \eqref{lnlqh} and \eqref{slq},
we propose the following simple ansatz
\begin{equation}
  \lnf = \frac{\ceffp}{4} \ln \left( t^2 \frac{\ell-t}{\ell+t} \right) + \text{const.}
  \label{lnlqfit}
\end{equation}
Note that one has only two fitting parameters, namely the prefactor $\ceffp$ as
well as the constant. Clearly, for the limiting cases $\lambda=0$ and $\lambda=1$,
one has to have $\ceffp=0$ and $\ceffp=c=1$, respectively.

\begin{figure}[H]
    \centering   
    \hspace*{-0.25cm}
    \subfloat{\includegraphics{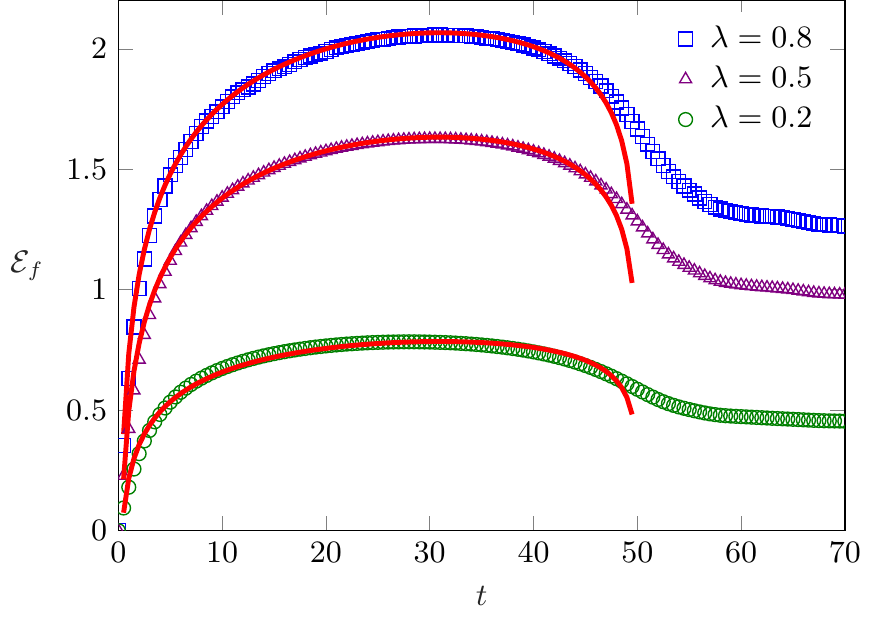}}
    \qquad
    \subfloat{\includegraphics{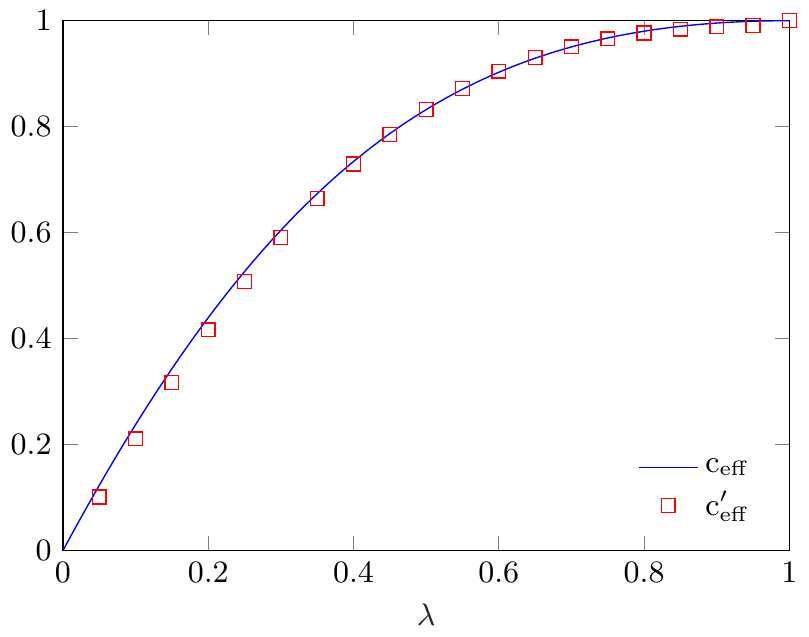}}
    \caption{Left: time evolution of the logarithmic negativity $\lnf$ after a quench from equal
    fillings, for $\ell=50$ and $2N = 600$. The red solid lines show the ansatz \eqref{lnlqfit}
    the data is fitted to. Right: comparison of the fitted values of $\ceffp$ to the ground-state
    effective central charge $\ceff$ from \eqref{ceff}}
    \label{fig:lnlqfit}
\end{figure}

For intermediate values of $\lambda$, we have determined $\ceffp$ by fitting the
ansatz \eqref{lnlqfit} to the numerically calculated $\lnf$ curves for a fixed interval
length $\ell=50$. The results are shown in Fig.~\ref{fig:lnlqfit}. On the left of the figure
the numerical data is shown together with the ansatz \eqref{lnlqfit}. On the right
hand side we plot the obtained values of $\ceffp$ as a function of $\lambda$, compared
to the equilibrium effective central charge $\ceff$. One can clearly see that the
two functions behave very similarly and one has $\ceffp \approx \ceff$.
Indeed, for larger values of $\lambda$ the agreement is almost perfect, however
the deviation increases for smaller defect strengths. This is also obvious from the
left of Fig.~\ref{fig:lnlqfit}, where the data for $\lambda=0.2$ shows already some
larger discrepancy compared to the fit function. 

%%%%%%%%%%%%%%%%%%%%%%%%%%%%%%%%%%%%%%%%%
%%%%%%%%%%%%%%%%%%%%%%%%%%%%%%%%%%%%%%%%%

Although the mismatch of the data might be due to finite-size effects, 
we observe essentially the same behaviour for larger $\ell=75$.
This rather suggests that the simplistic ansatz \eqref{lnlqfit} is probably not the
exact leading functional form of $\lnf$.
%In fact, we have also tried a more complicated ansatz, where the various logarithmic terms,
%with arguments $t$, $\ell-t$ and $\ell+t$, respectively, were assigned different prefactors.
%Unfortunately, however, the fits turn out to be very unstable against changing the fitting
%interval, making the results unreliable. This is not very surprising, since one then has a
%four-parameter function to fit, with very slowly varying functions.
In fact, we have also tried a more complicated three-parameter ansatz, assigning two different 
prefactors to the logarithmic terms with arguments $t$ and $(\ell-t)/(\ell+t)$. Unfortunately, 
however, the fits turn out to be very unstable against changing the fitting interval, 
making the results unreliable. Thus we conclude that, without some additional insights 
(e.g. from CFT calculations) on the structure of $\lnf$, extracting the proper time-dependece
from numerical calculations is a very hard task.

We have also compared the behaviour of $\lnf$ and $\mi$ directly. One finds that
the curves almost exactly overlap until roughly $t \approx \ell$, with no visible deviations.
Around $t = \ell$ there is only a slight deviation between the two functions, which,
however diminishes again for $t \gg \ell$. Indeed, both quantities are expected to 
converge towards their ground-state values asymptotically, where the deviation was
already found to be tiny, see Fig.~\ref{fig:gs}.
Therefore, for better visibility of the data, we have not included $\mi$
in the left of Fig.~\ref{fig:lnlqfit}.

\section{Quench from unequal fillings\label{sec:quench2}}

We now study initial states with unequal fillings, where the behaviour of the
entanglement negativity turns out to be qualitatively different from the unbiased
case discussed above. In most of our calculations we shall actually consider
the maximally biased case, $n_l=1$ and $n_r=0$, while at the end of the section
we show that the generalization to arbitrary fillings is straightforward.
Similarly to the unbiased case, we first discuss the negativity evolution for
a homogeneous chain $(\lambda=1)$, where results can also be obtained
via CFT techniques.

\subsection{Homogeneous chain}

This case is also known as the domain-wall quench, due to the form of the
initial state in the spin-chain equivalent of the hopping model. 
Here one can work directly in the thermodynamic limit, $N\to\infty$, where 
the correlation matrix is known exactly and has the simple form \cite{eisler2013fullcounting}
\begin{equation}
  C_{mn}(t) = \frac{i^{n-m}\, t}{2(m-n)} \left[ J_{m-1}(t)J_n(t) - J_m(t)J_{n-1}(t) \right] ,
  \label{eq: DW_tdlim}
\end{equation}
where $J_m(t)$ is the Bessel function of order $m$. Remarkably, the correlation
matrix for the domain-wall quench is unitarily equivalent to the one in a static ground-state problem,
namely a hopping chain with a linear chemical potential \cite{EIP09}, with the time $t$
playing the role of the characteristic length scale of the potential. This can actually be shown to be a
particular example of a more general mapping, known as emergent eigenstate solution \cite{VIR17}.
Consequently, the entanglement properties in the dynamical and static problems are identical.

The entanglement entropy for a biased hopping chain was studied in
\cite{EIP09,EP14,alba2014entanglement,gruber2019magnetization} between
two parts of the chain, with the cut located somewhere along the emerging front.
The growth was found to be logarithmic, however with a different prefactor as for the
local (unbiased) quench. The analytical understanding of the results for $n_l=1$ and $n_r=0$
came afterwards, when the method of curved-space CFT was developed \cite{DSVC17}.
The basic idea is that certain inhomogeneous free-fermion problems can be treated by first
mapping the problem to a CFT in a curved background metric. The entanglement entropies can be
calculated by applying standard replica-trick methods \cite{CC09} for the curved-space Dirac
fermion theory (see also \cite{BDS19} for recent results on inhomogeneous Luttinger liquids).
This makes it possible to extract the entropy analytically for a half-chain, or even generalize
the calculations to a finite segment within the front region \cite{DSVC17}.
The mutual information $\mi$ in our setup can be obtained immediately from the latter result.

Furthermore, it is possible to combine the curved-space technique with the CFT
approach to the negativity \cite{CCT12,CCT13}. The calculation is straightforward but somewhat
lengthy, thus we present it in Appendix \ref{app:cft}. As a result, we obtain for both the negativity
and the mutual information
\begin{equation}
  \lnf = \mi= \frac{1}{4} \ln \left[ \, t \, f\left(\ell/t\right) \right] + c_1, \qquad
  f(\xi) = \begin{cases}
  1 & t < \ell \\
  \frac{\left(1-\sqrt{1-\xi^2} \right)^2}{\xi^3} & t > \ell
  \end{cases}
  \label{lndwhom}
\end{equation}
Note that, apart from an explicit factor of $t$ which gives the logarithmic growth of the negativity,
\eqref{lndwhom} depends only on a scaling function of the variable $\xi=\ell/t$.
It should be stressed that the calculation in Appendix \ref{app:cft} refers only to $\xi<1$, as the
curved-space CFT is able to describe only the front region with nontrivial fermionic density.
However, the $\xi>1$ result can be obtained by using continuity and some simple arguments.
Indeed, for $\ell > t$, the segments include parts of the chain outside the front, where the density
is either zero or one. Clearly, these pieces do not contribute to the entanglement at all, which
is thus given by the result for $\ell=t$, i.e. by the limit $\xi \to 1$.

\begin{figure}[H]
  \centering
  \includegraphics{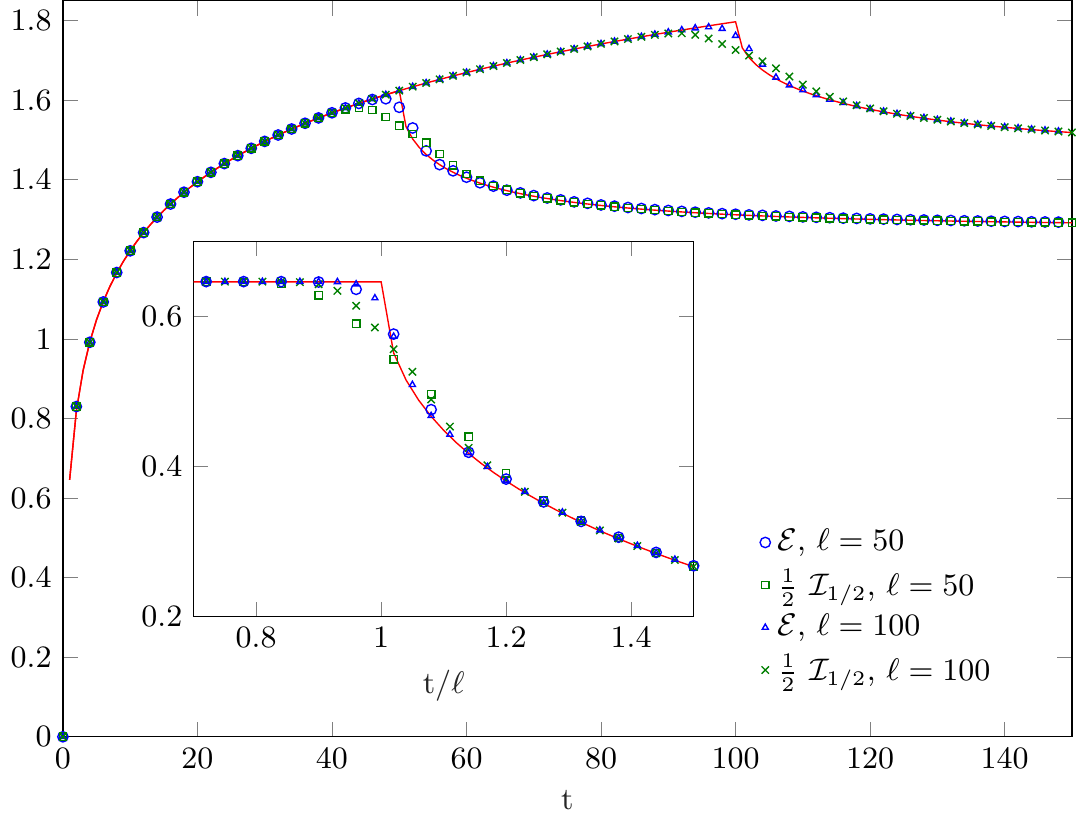}
  \caption{Logarithmic negativity $\lnf$ and mutual information $\mi$ after the domain wall quench
  in a homogeneous chain ($\lambda=1$), compared to the CFT prediction \eqref{lndwhom} (red solid lines)
  for two different segment sizes. The inset shows the scaled data around $t = \ell$, after substracting $\frac{1}{4} \ln(t)$.}
  \label{fig: DW_hom}
\end{figure}

The CFT results in \eqref{lndwhom} are compared to our numerical calculations in Fig.~\ref{fig: DW_hom}
with a very good agreement. The constant $c_1 \approx 0.646$ has been obtained by fitting the
data for $\lnf$ in the regime $t < \ell$. The only visible deviations are around $t = \ell$, i.e. when
the boundaries of the segments are close to the edges of the front. Indeed, the front is known to have
a nontrivial scaling behaviour around the edge \cite{eisler2013fullcounting,VSDH15,Fagotti17}, characterized by the scale $t^{1/3}$,
which is not resolved by the CFT treatment. Nevertheless, when plotted against $t/\ell =\xi^{-1}$
and after subtracting $1/4 \ln (t)$, the lattice data for increasing $\ell$ converge smoothly towards
the CFT result as shown in the inset of Fig.~\ref{fig: DW_hom}. Note also that the $t \to \infty$
behaviour can be obtained by expanding $f(\xi)\approx \xi$/4 for $\xi \to 0$, such that
the steady state is characterized by $\lnf = 1/4 \ln \ell + \const$.

%%%%%%%%%%%%%%%%%%%%%%%%%%%%%%%%%%%%%%%%%
%%%%%%%%%%%%%%%%%%%%%%%%%%%%%%%%%%%%%%%%%

\subsection{Chain with a defect}

The time evolution of entanglement across a defect turns out to be qualitatively different
\cite{eisler2012onentanglement}.
Indeed, an imperfect transmission between the half-chains gives rise to scattering,
i.e. the single-particle modes are partially transmitted and reflected with probabilities
$T_q$ and $R_q=1-T_q$, respectively. For a weak hopping defect parametrized by
$\lambda=\ee^{-\nu}$, the transmission coefficient is given by
  \begin{equation}
    T_q = \frac{\sin^2(q)}{\cosh^2(\nu) - \cos^2(q)} \, .
    \label{Tq}
  \end{equation}
The transmitted and reflected particles become entangled in the wavefunction,
and the contribution of such a pair in the $\alpha=1/2$ R\'enyi entropy is $s_{1/2}(T_q)$, where
\begin{equation}
    s_{1/2}(x) = 2 \ln\left( x^{1/2} + (1-x)^{1/2} \right)
\end{equation}
is the density of the R\'enyi entropy, c.f. Eq. \eqref{eq: S_alpha}.
Now, due to the density bias in the initial state, there is a constant influx of
particles and consequently a steady generation of entanglement at the defect.
For a half-chain, $\ell=N$, at maximum bias and in the limit $N\to\infty$, this was found
to give, to leading order, a linear growth of entanglement \cite{eisler2012onentanglement}
\begin{equation}
  S_{1/2}(\rho_{A_1}) = t \int_0^\pi \frac{\dd q}{2\pi} \, v_q \, s_{1/2}(T_q) \, ,
  \label{sdw}
\end{equation}
where $v_q = \sin(q)$ is the single-particle group velocity and the integral
is taken over all modes with $v_q>0$.

This simple semiclassical picture of entanglement production, based on the propagation of
entangled pairs of quasiparticles, bears a strong similarity to global quenches \cite{CC05,AC17}.
One should stress, however, that here the pairs are created solely at the defect site
but steadily in time, in contrast to a global quench where pairs are created only
at $t=0$ but homogeneously along the chain. Nevertheless, the continuity argument
$\lnf \simeq \mi$ can be applied the very same way as for the global quench \cite{alba2019quantuminfo}.
Indeed, due to the strictly local production of entanglement at the common boundary
of the segments, the only effect of their finite size is to cut off the growth of
the negativity once the distance travelled by a given mode $v_q t$ exceeds the segment size $\ell$.
This leads to the semiclassical expression
\begin{equation}
 \lneg_{sc} = \int_{0}^{\pi} \frac{\dd q}{2\pi} \min{(v_qt,\ell)} \, s_{1/2}(T_q) \, .
%  {q_{F,R}}^{q_{F,L}}
  \label{lndwsc}
\end{equation}
%
%Note that, in this simple picture, the role of $S_{1/2}(\rho_A)$ in the mutual information
%is simply to remove the (ground-state) entanglement contribution from the homogeneous
%boundaries. The entangled pairs do not contribute to this piece, as they are either
%both inside or both outside the composite interval $A$.

\begin{figure}[H]
  \centering
  \includegraphics{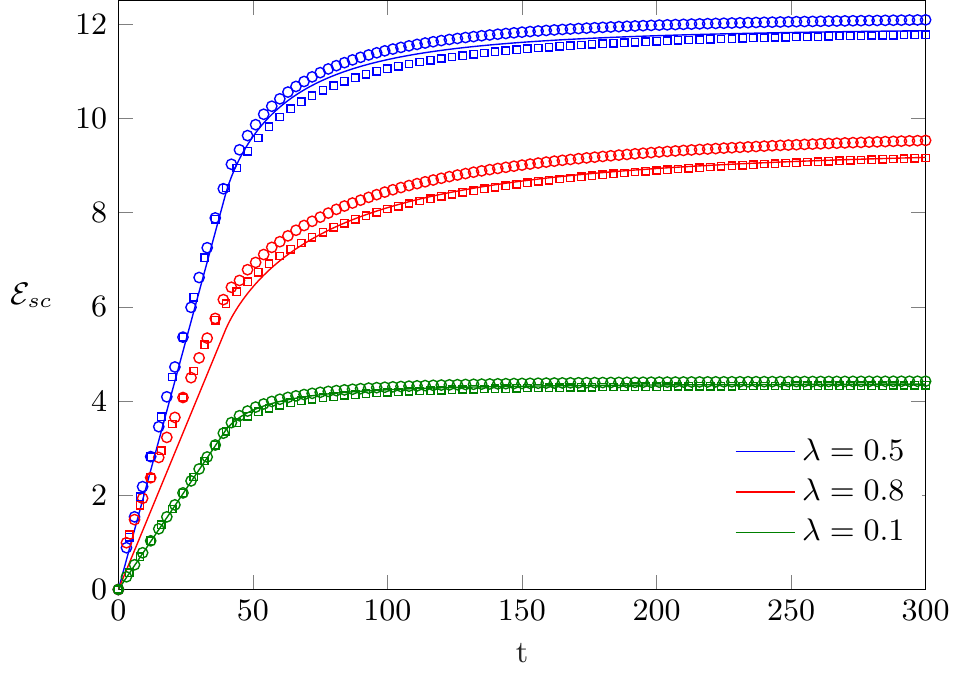} 
  \caption{Entanglement negativity $\lnf$ (circles) and mutual information $\mi$ (squares) after a quench from
  unequal fillings $n_l=1$ and $n_r=0$, compared to the semiclassical ansatz \eqref{lndwsc} (lines) for
  $\ell=40$ and different values of $\lambda$.}
  \label{fig: DW_sc}
\end{figure}

To test the validity of our ansatz $\lneg_{sc}$, in Fig.~\ref{fig: DW_sc} we plot the integrals
\eqref{lndwsc} for various $\lambda$ and compare them to the numerical data
for $\lnf$ and $\mi$ obtained for a chain of length $2N = 500$
from the correlation-matrix method.
One can see that the semiclassical picture provides a rather good description of
the data to leading order, there are, however, still some sizeable corrections which
tend to diminish for smaller values of $\lambda$.
The quantities $\lnf$ and $\mi$ perfectly overlap in the regime $t < \ell$ of linear growth,
where the result is identical to the one for the half-chain \eqref{sdw}, as there is no
contribution to the entanglement  from the outer boundaries of the segments.
Interestingly, however, there is a clearly visible splitting for $t>\ell$, after the front has
traversed the segments.

%%%%%%%%%%%%%%%%%%%%%%%%%%%%%%%%%%%%%%%%%
%%%%%%%%%%%%%%%%%%%%%%%%%%%%%%%%%%%%%%%%%

\begin{figure}[H]
  \centering
    \hspace*{-0.25cm}
    \subfloat{\includegraphics{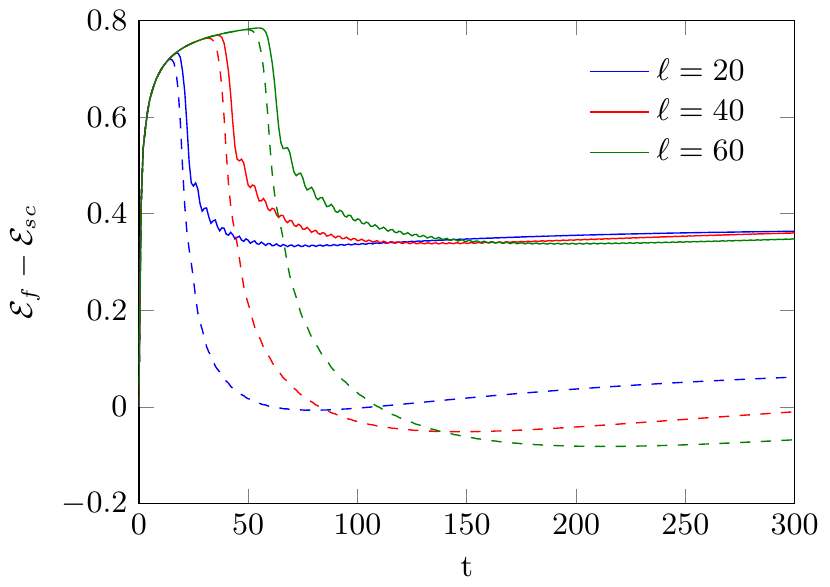}}    
    \subfloat{\includegraphics{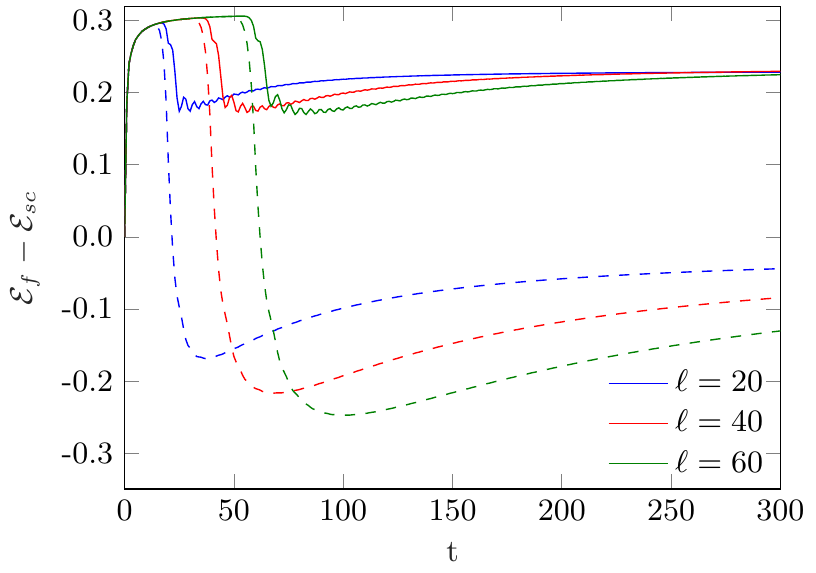}} 
  \caption{Deviations of $\lnf$ (solid lines) and $\mi$ (dashed lines) from the semiclassical ansatz \eqref{lndwsc}
  for $\lambda = 0.8$ (left) and $\lambda = 0.5$ (right) and various segment sizes $\ell$.}
  \label{fig: DW_sub}
\end{figure}

In order to better understand the corrections beyond the semiclassical picture,
in Fig.~\ref{fig: DW_sub} we have subtracted $\lneg_{sc}$ from the data, shown for
the two larger values of $\lambda$ and several segment sizes. Similarly to $\lneg_{sc}$,
the deviation also shows different behaviour in the regimes $t<\ell$ and $t>\ell$.
Until roughly $t\approx\ell$, one observes a steady growth which becomes slower
for larger defect strengths. A closer inspection shows that this subleading growth
is actually slower than logarithmic for all the values $\lambda \ne 1$ we have checked.
When the front crosses the segment boundary, one has a sharp drop in all of the curves,
which is then followed again by a very slow increase. Note that the splitting of the
$\lnf$ and $\mi$ curves is even more apparent in Fig.~\ref{fig: DW_sub}.
However, due to the very slow variation of the data, it is hard to draw a firm
conclusion about the asymptotic behaviour, despite the relatively large times
considered in the calculations.

%%%%%%%%%%%%%%%%%%%%%%%%%%%%%%%%%%%%%%%%%
%%%%%%%%%%%%%%%%%%%%%%%%%%%%%%%%%%%%%%%%%

\begin{figure}[H]
  \centering
    \hspace*{-0.25cm}
    %\scalebox{1}{\input{./PLOTS/TEX/neg_tdlim.tex}}\\
    %\subfloat{\includegraphics{./PLOTS/PDF/Imut_tdlim.pdf}}\\
    \subfloat{\includegraphics{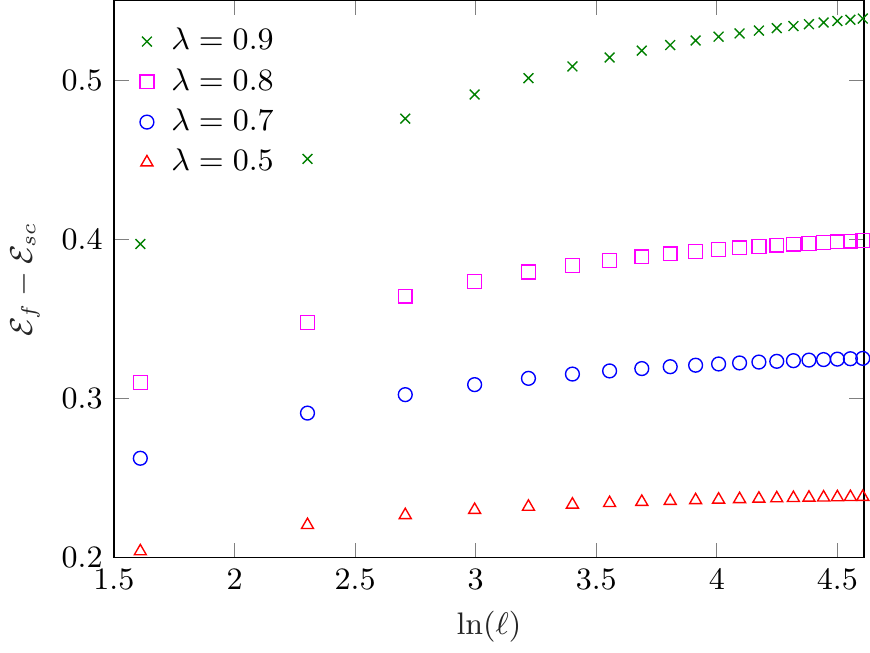}}
    \
    \subfloat{\includegraphics{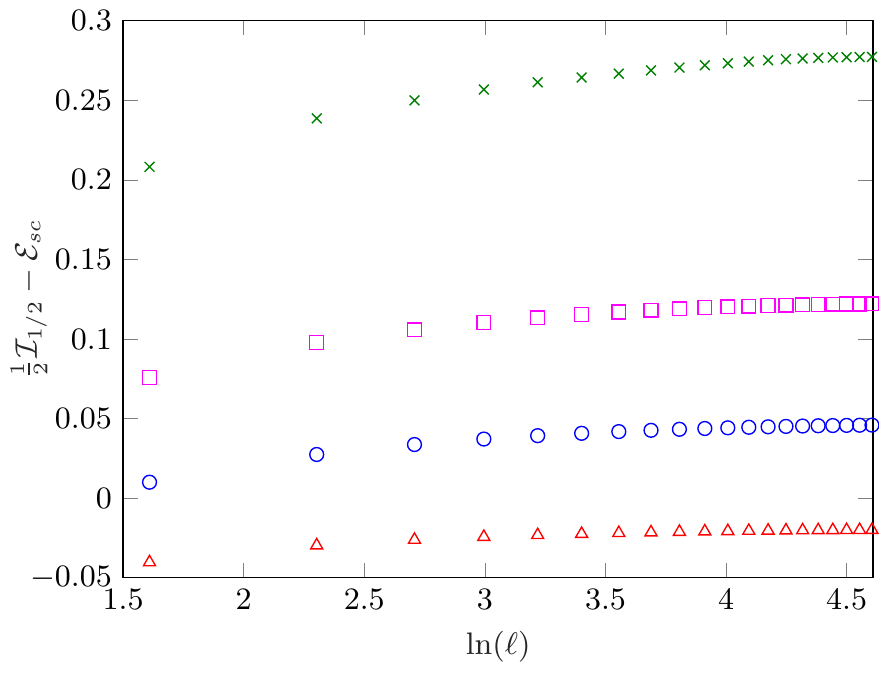}}
  \caption{Steady-state negativity (left) and mutual information (right), as calculated from the correlation matrix
  \eqref{cdwness} after subtracting the extensive semiclassical contribution \eqref{lndwsc}. The data is plotted
  against $\ln (\ell)$ for various defect strengths $\lambda$. Note the different vertical scales.}
  \label{fig: DW_ness}
\end{figure}

The steady state after the quench across the defect can actually be captured directly.
Indeed, the elements of the correlation matrix in \eqref{eq: C(t)} have a well defined limit \cite{LSP19}
\eq{
\tilde C_{m,n} = \lim_{t \to \infty}\lim_{N \to \infty} C_{m,n}(t) \, ,
\label{cdwness}}
with the explicit formulas collected in Appendix \ref{app:corr}.
These can be used to evaluate the subleading scaling of the steady-state negativity
and mutual information, i.e. after subtracting the extensive contribution
$\lim_{t\to\infty}\lneg_{sc}$ that follows from the semiclassical description \eqref{lndwsc}.
The results are shown in Fig.~\ref{fig: DW_ness}, with both the data for $\lnf$ (left)
as well as $\mi$ (right) plotted against $\ln (\ell)$. Rather clearly, the subleading terms
in the steady state are different for the two quantities and the scaling in $\ell$ is
slower than logarithmic. This is also supported by the form of the steady-state
correlation matrix \eqref{c0ness} on a given side of the defect, which is a Toeplitz
matrix with a symbol given by $T_q$ for $q>0$ and zero otherwise.
While this yields immediately the extensive part of the negativity \eqref{lndwsc},
one has also $T_q \to 0$ for $q \to 0$ and thus no jump singularity is present.
Nevertheless, the symbol is still nonanalytic and shows a very sharp increase
around $q=0$ as $\lambda \to 1$. Thus a weaker than logarithmic growth of the
subleading term, although unlikely from the numerics, cannot be excluded.

%Note that for $\lambda = 0.5$, the mutual information subtracted by the linear
%contribution (see Fig.~\ref{fig: DW_ness} (right)) is negative.

%%%%%%%%%%%%%%%%%%%%%%%%%%%%%%%%%%%%%%%%%
%%%%%%%%%%%%%%%%%%%%%%%%%%%%%%%%%%%%%%%%%

\begin{figure}[H]
  \centering
    %\hspace*{-0.5cm}
    \subfloat{\includegraphics{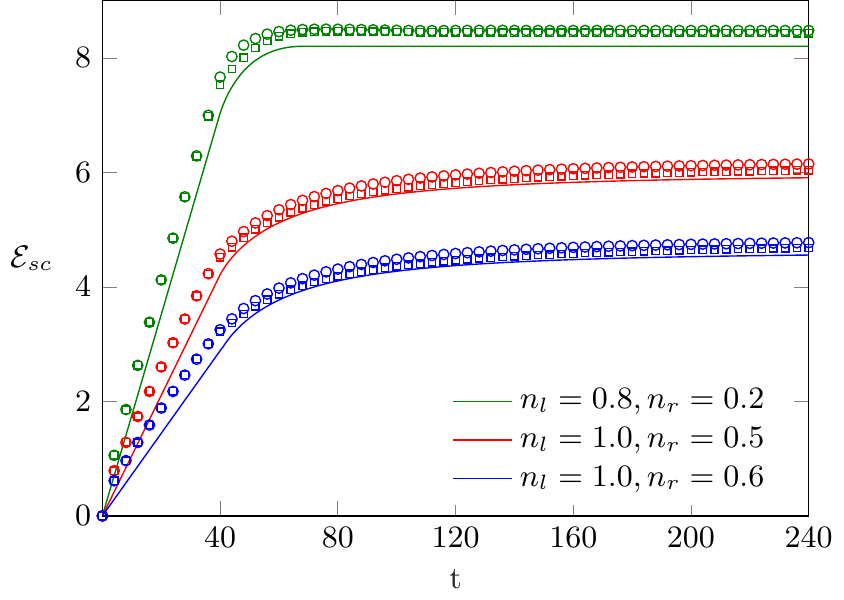}}
    \subfloat{\includegraphics{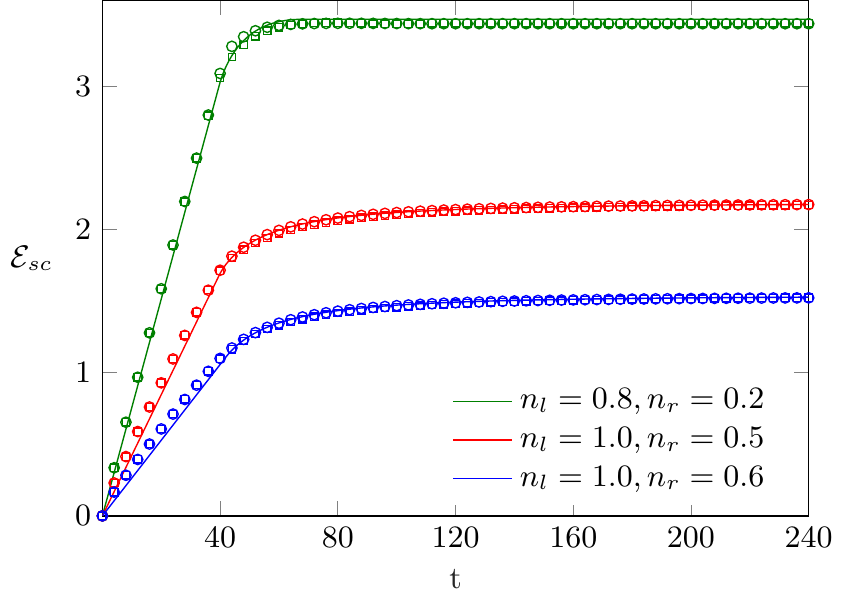}}
    \caption{Entanglement negativity $\lnf$ (circles) and mutual information $\mi$ (squares) 
   after a quench from various unequal fillings, compared to the semiclassical ansatz \eqref{lndwsc2} (lines)
   for $\ell=40$ and $\lambda = 0.5$ (left) resp. $\lambda = 0.1$ (right).}
  \label{fig: diff_fill}
\end{figure}

Finally, we briefly consider the case of arbitrary fillings with $n_l > n_r$.
The straightforward generalization of the semiclassical ansatz reads
\begin{equation}
 \lneg_{sc} = \int\displaylimits_{q_{F,r}}^{q_{F,l}} \frac{\dd q}{2\pi} \min{(v_qt,\ell)} \, s_{1/2}(T_q) \, ,
  \label{lndwsc2}
\end{equation}
where the integral is carried out only between the Fermi wavenumbers
$q_{F,\sigma}=\pi n_{\sigma}$. In other words, one has to consider only the
contributions from the uncompensated fermionic modes, that can propagate
from the left to the right half-chain. The resulting curves are shown in 
Fig.~\ref{fig: diff_fill}, for two different $\lambda$ and various fillings,
together with the numerically calculated $\lnf$ and $\mi$ in a chain of size $2N=500$.
As expected, the plots are very similar to the one in Fig.~\ref{fig: DW_sc},
with the deviations from the semiclassical prediction decreasing for smaller $\lambda$.
Due to the similar qualitative behaviour, a detailed analysis of the subleading terms
is not presented for this case.

%%%%%%%%%%%%%%%%%%%%%%%%%%%%%%%%%%%%%%%%%
%%%%%%%%%%%%%%%%%%%%%%%%%%%%%%%%%%%%%%%%%

\section{Quench in the XXZ chain with a defect\label{sec:xxz}}

Finally we are considering a quench in the XXZ spin chain, given by the Hamiltonian
\begin{equation}
  \hat{H}_{XXZ} = \sum_{j=-N+1}^{N-1}
  \left[ J_j \left(S_j^x S_{j+1}^x + S_j^y S_{j+1}^y \right) + \Delta S_j^z S_{j+1}^z \right] ,
  \label{hxxz}
\end{equation}
where $S_j^\alpha$ are spin-$1/2$ operators, $\Delta$ is the anisotropy parameter,
and the XX-coupling is given by
\begin{equation}
    J_j = \begin{cases}
           1 &\text{if}\ j \neq 0 \, , \\
           \lambda \, \Theta(t) &\text{if}\ j = 0 \, .
          \end{cases}
\end{equation}
Here $\Theta(t)$ is the Heaviside step function, in other words, the quench consists of
simply joining two decoupled XXZ half-chains at time $t=0$.
Applying a Jordan-Wigner transformation, the Hamiltonian \eqref{hxxz} can be mapped
into a chain of interacting fermions and the setup becomes exactly the same
as the one depicted in Fig.~\ref{fig: setup} for free fermions without a density bias.

However, it turns out that the negativity depends on the choice of basis and is not
equivalent in the fermion or spin representation. 
Indeed, using spin variables, one has to apply the conventional definition of
the logarithmic negativity via the partial transpose of the density matrix \cite{vidal2002computable}
\begin{equation}
  \lns = \ln \norm{\rho_A^{T_{2}}}_1
  \label{lns} \, .
\end{equation}
Here the partial transpose is taken with respect to subsystem $A_2$,
defined by its matrix elements as
\begin{equation}
  \bra{e^{(1)}_i,e^{(2)}_j} \rho_A^{T_{2}} \ket{e^{(1)}_k,e^{(2)}_l} =
  \bra{e^{(1)}_i,e^{(2)}_l} \rho_A \ket{e^{(1)}_k,e^{(2)}_j},
  \label{pt}
\end{equation}
where $\ket{e^{(1)}_i}$ and $\ket{e^{(2)}_j}$ denote orthonormal bases on the
Hilbert spaces pertaining to segments $A_1$ and $A_2$.

Clearly, since we are now faced with a non-Gaussian problem, we have to compute
the negativity $\lns$ via density-matrix renormalization group (DMRG) \cite{schollwoeck2011density,white1992DMRG,white1993DMRG} methods.
In particular, the time evolution $\ket{\psi(t)} = e^{-i \hat H_{XXZ} t} \ket{\psi(0)}$ after the quench is first performed with time-dependent DMRG
(tDMRG) simulations \cite{daley2004tMDRG,white2004tDMRG}, which give access to the reduced density matrix $\rho_A$
in a matrix product state (MPS) representation \cite{itensor}.
The partial transpose and the corresponding logarithmic negativity can then be calculated
using the method of Ref.~\cite{ruggiero2016entanglement}, which is briefly reviewed in the
following subsection. 

\subsection{Negativity for matrix product states}

Let us consider the time-evolved state after the quench
$\ket{\psi(t)} = e^{-i \hat H_{XXZ} t} \ket{\psi(0)}$ and its MPS representation
\begin{equation}
  \ket{\psi(t)} = \sum_{\sigma_1 \dots \sigma_{2N}}
  T_{\nu_1}^{\sigma_1} T_{\nu_1 \nu_2}^{\sigma_2} \dots T_{\nu_{2N-1}}^{\sigma_{2N}}
  \ket{\sigma_1 \dots \sigma_{2N}},
  \label{eq: psi_MPS}
\end{equation}
where $T_{\nu_{i-1} \nu_{i}}^{\sigma_{i}}$ denotes the tensor on site $i$ with bond indices
$\nu_{i-1}$ and $\nu_{i}$ and the physical index $\sigma_{i}=0,1$.
In Eq.~\eqref{eq: psi_MPS} and all the following equations, we assume summation over all repeated
bond indices $\nu_i$ implicitly, and indicate only summations over the physical indices for better readability.

\begin{figure}[H]
    \centering
    \includegraphics{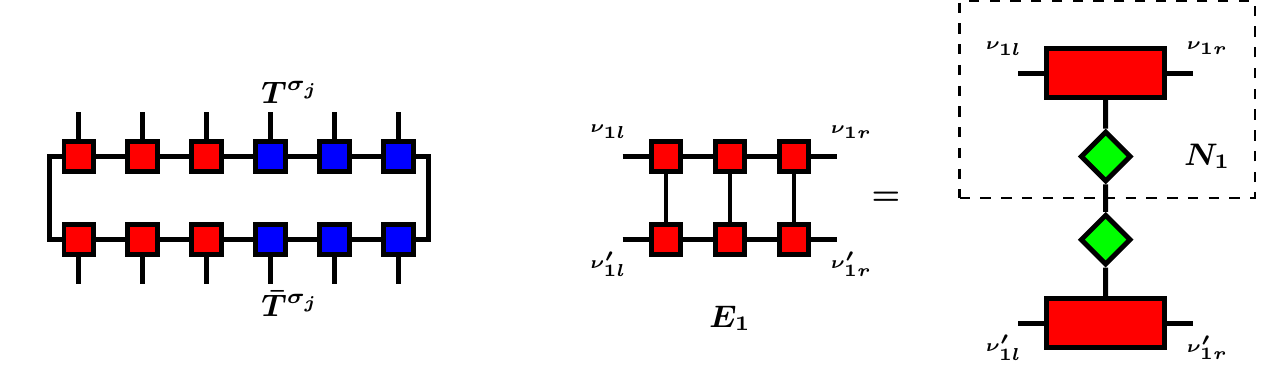}
    \caption{Graphical representation of the reduced density matrix $\rho_{A}$ (left),
    the transfer matrix $E_1$ corresponding to the segment $A_1$ (middle),
    and its singular value decomposition (right). The boxes represent tensors
    with the contractions indicated by the links between them.}
  \label{fig: E_matrices}
\end{figure}

Our goal is to calculate the negativity for the geometry depicted in Fig.~\ref{fig: setup},
i.e. between two adjacent segments $A_1$ and $A_2$ with $A=A_1 \cup A_2$.
The main step is to construct the reduced density matrix $\rho_A$, which is shown graphically
on the left of Fig.~\ref{fig: E_matrices}, after tracing out over the environment $B$.
The squares in different colors depict the tensors
belonging to either subsystems $A_1$ or $A_2$, c.f. Fig.~\ref{fig: setup}.
One can now introduce new basis states in the Hilbert spaces of the two intervals $A_1$ and $A_2$ as
\begin{equation}
    \ket{w^{(1)}_{\nu_{1l},\nu_{1r}}} = \sum_{\{\sigma_i\}}
    \prod_{i \in A_1} T_{\nu_{i-1} \nu_{i}}^{\sigma_{i}} \ket{\sigma_i} ,
    \qquad
    \ket{w^{(2)}_{\nu_{2l},\nu_{2r}}} = \sum_{\{\sigma_i\}}
    \prod_{i \in A_2} T_{\nu_{i-1} \nu_{i}}^{\sigma_{i}} \ket{\sigma_i} ,
    \label{w12}
\end{equation}
where the index pairs $\nu_{1l},\nu_{1r}$ and $\nu_{2l},\nu_{2r}$ indicate the uncontracted
left- and rightmost bond indices for each block.
Using these basis states, we can eventually write the reduced density matrix as 
\begin{equation}
    \rho_{A} =
    \delta_{\nu_{1l},\nu'_{1l}} \delta_{\nu_{2r},\nu'_{2r}}
    \delta_{\nu_{1r},\nu_{2l}} \delta_{\nu'_{1r},\nu'_{2l}}
    \ket{w^{(1)}_{\nu_{1l},\nu_{1r}}} \bra{w^{(1)}_{\nu'_{1l},\nu'_{1r}}} \otimes
    \ket{w^{(2)}_{\nu_{2l},\nu_{2r}}} \bra{w^{(2)}_{\nu'_{2l},\nu'_{2r}}},
    \label{eq: rho_basis1}
\end{equation}
where the delta functions carry out the contractions of the remaining bond indices,
as visualized in the left of Fig.~\ref{fig: E_matrices}.

%%%%%%%%%%%%%%%%%%%%%%%%%%%%%%%%%%%%%%%%%%%
%%%%%%%%%%%%%%%%%%%%%%%%%%%%%%%%%%%%%%%%%%%

The representation \eqref{eq: rho_basis1} yields a decomposition of $\rho_A$ on the
two subspaces corresponding to $A_1$ and $A_2$.
However, the main problem is that the choice of basis in \eqref{w12} is not orthogonal.
Indeed, the overlaps between these states are given by the so-called transfer matrices
\begin{equation}
    \braket{w^{(1)}_{\nu_{1l},\nu_{1r}}|w^{(1)}_{\nu'_{1l},\nu'_{1r}}} = \left[E_1\right]^{\nu_{1l},\nu_{1r}}_{\nu'_{1l},\nu'_{1r}},
    \qquad
    {\braket{w^{(2)}_{\nu_{2l},\nu_{2r}}|w^{(2)}_{\nu'_{2l},\nu'_{2r}}}} = \left[E_2\right]^{\nu_{2l},\nu_{2r}}_{\nu'_{2l},\nu'_{2r}},
\end{equation}
that are obtained by contracting all the tensors with their complex conjugates
via their physical indices within the respective segment
\begin{equation}
    \left[E_1\right]^{\nu_{1l},\nu_{1r}}_{\nu'_{1l},\nu'_{1r}} = \prod_{i \in A_1} \sum_{\sigma_i} 
    T_{\nu_{i-1} \nu_{i}}^{\sigma_i} \bar{T}_{\nu'_{i-1} \nu'_{i}}^{\sigma_i} \, , \qquad 
    \left[E_2\right]^{\nu_{2l},\nu_{2r}}_{\nu'_{2l},\nu'_{2r}} = \prod_{i \in A_2} \sum_{\sigma_i} 
    T_{\nu_{i-1} \nu_{i}}^{\sigma_i} \bar{T}_{\nu'_{i-1} \nu'_{i}}^{\sigma_i} \, .
    \label{E12}
\end{equation}
These objects are thus four-index tensors, corresponding to the uncontracted
left- and rightmost bond indices, see the middle panel of Fig.~\ref{fig: E_matrices}
for a graphical representation of $E_1$.

In order to obtain an orthogonal basis, one has to perform a singular value
decomposition (SVD) of the transfer matrices $E_1=U_1D_1V_1^\dagger$
and $E_2=U_2D_2V_2^\dagger$. This amounts to introducing a basis change via
 \begin{equation}
   \ket{w^{(1)}_{\nu_{1l},\nu_{1r}}} = \sum_m \left[ N_1 \right]^m_{\nu_{1l},\nu_{1r}} \ket{v_m^{(1)}},
   \qquad
   \ket{w^{(2)}_{\nu_{2l},\nu_{2r}}} = \sum_n \left[ N_2 \right]^n_{\nu_{2l},\nu_{2r}} \ket{v_n^{(2)}},
   \label{w12b}
\end{equation}
where $N_1 = U_1D_1^{1/2}$ and $N_2 = U_2D_2^{1/2}$, and the new indices
$m$ and $n$ correspond to the singular values contained in the diagonal matrices
$D_1$ and $D_2$. The pictorial representation of the SVD for $E_1$ is shown
on the right of Fig.~\ref{fig: E_matrices}, where $D_1^{1/2}$ is depicted by the
green rhombi. 
Inserting \eqref{w12b} into \eqref{eq: rho_basis1}, one immediately obtains
the matrix elements of the reduced density matrix
\begin{equation}
  \left[ \rho_{A} \right]^{m,m'}_{n,n'} = 
  \left[N_1\right]^m_{\nu_{1l},\nu_{2l}} \left[\bar{N}_1\right]^{m'}_{\nu_{1l},\nu'_{2l}}
%  \delta_{\nu_{1r},\nu_{2l}}\delta_{\nu'_{1r},\nu'_{2l}} 
  \left[N_2\right]^n_{\nu_{2l},\nu_{2r}} \left[\bar{N}_2\right]^{n'}_{\nu'_{2l},\nu_{2r}}
%  \ket{v_m^{(1)}} \otimes \ket{v_n^{(2)}} \bra{v_m'^{(1)}} \otimes \bra{v_n'^{(2)}}
  \label{eq: rho_basis2}
\end{equation}
expressed in the orthogonal bases $\ket{v_m^{(1)}}$ and $\ket{v_n^{(2)}}$.
Note that \eqref{eq: rho_basis2} is now exactly in the form required to carry out
the partial transposition according to \eqref{pt}. Indeed, the index pairs $m, m'$
and $n, n'$ correspond to the intervals $A_1$ and $A_2$, respectively.
Therefore, the matrix elements of $\rho^{T_{2}}_{A}$ can simply be obtained
by exchanging $n$ and $n'$. Finally, the logarithmic negativity $\lns$ in
Eq.~\ref{lns} can be calculated via an explicit diagonalization of $\rho^{T_{2}}_{A}$.

Regarding the computational effort, one has to stress that the cost of constructing the
transfer matrices in \eqref{E12} scales as $\mathcal{O}(\chi_{max}^6)$ with the maximum bond
dimension $\chi_{max}$ of the MPS. This, however, grows with the time evolution where
we set the requirement $\epsilon \sim 10^{-8} - 10^{-9}$ for the truncated weight.
For a feasible computation of the transfer matrices, we truncated back the bond dimension
to $\chi_{max} = 300$. The range of each index $m,m'$ as well as $n,n'$ in the representation
\eqref{eq: rho_basis2} is then bounded by $\chi^2_{max}$, which is still too large for a
tractable calculation. However, since the singular values of $E_1$ and $E_2$ decay
rapidly, one can apply a truncation after the SVD which we set to $\chi'_{max}=80$.
All in all, the evaluation of entanglement negativity is computationally much more
demanding than that of the entropy, severely limiting the attainable segment
sizes and simulation times in our numerics.

\subsection{Numerical results}

The methods outlined in the previous subsection are now used to evaluate
the time evolution of the logarithmic negativity across a defect in the XXZ chain.
We focus exclusively on the unbiased case, as the bias induces a much more
rapid growth of entanglement, which makes the DMRG calculations very demanding.
We first present the results for the XX chain, which is just the special case of
$\Delta = 0$ in \eqref{hxxz}. Note that, even though the XX chain with a defect
is exactly mapped into the fermionic Hamiltonian in Eq.~\eqref{eq: H} via a
Jordan-Wigner transformation, the negativity $\lns$ calculated for the spin chain
in \eqref{lns} is not equivalent to the fermionic one $\lnf$ defined in \eqref{lnf}.
Indeed, it has been shown in \cite{eisler2015onthept} that the partial transposition
in the spin basis yields a linear combination of two fermionic Gaussian operators
\begin{equation}
  \rho_A^{T_{2}} = \frac{1-i}{2} O_+ + \frac{1+i}{2} O_- \, ,
\end{equation}
where $O_+=\rho_A^{R_2}$ is the operator obtained in \eqref{ptr} by partial time
reversal of the fermionic degrees of freedom and $O_-=O_+^\dag$. Since in general
the operators $O_+$ and $O_-$ do not commute, one has no access to the
spectrum of $\rho_A^{T_{2}}$ and hence to $\lns$ via simple covariance-matrix techniques.
Nevertheless, the spin-chain negativity can be shown to be upper-bounded by the
fermionic one as \cite{herzog2016estimation,eisert2018entanglement}
\eq{
\lns \le \lnf + \ln \sqrt{2} \, .
\label{ub}
}

\begin{figure}[H]
  \centering
  \includegraphics{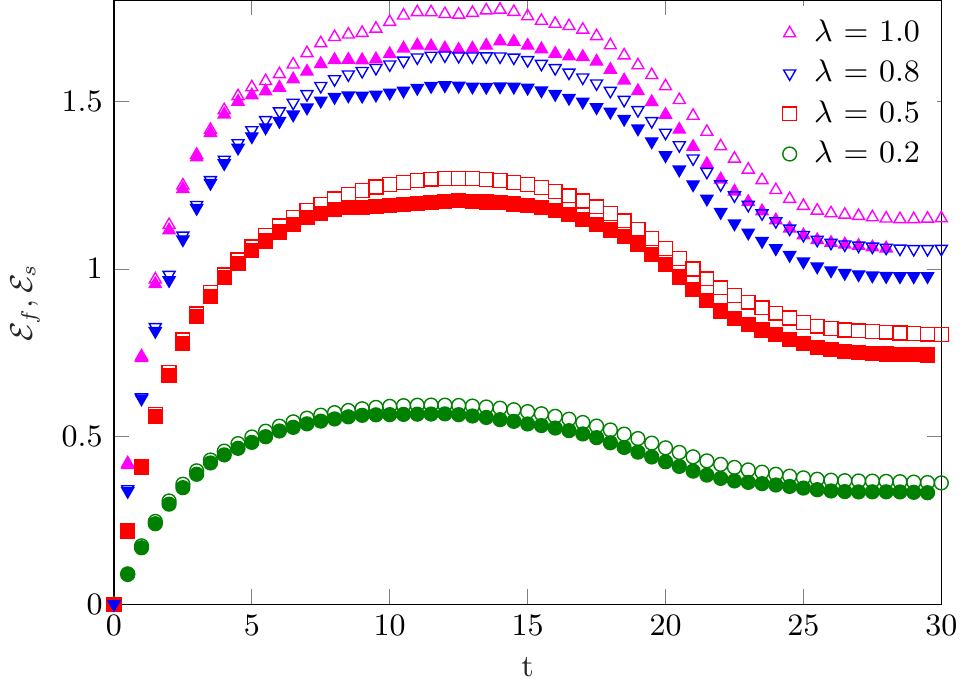}
  \caption{Time evolution of the entanglement negativity $\lns$ (full symbols) for two
  adjacent segments of size $\ell = 20$ across a defect of various strengths $\lambda$
  in an XX chain with $2N = 100$. The data for the fermionic negativity $\lnf$
  (empty symbols) in the analogous quench of the hopping chain with a defect
  is shown for comparison.}
  \label{fig: XX}
\end{figure}

The time evolution of $\lns$ obtained from tDMRG simulations are shown by the
full symbols in Fig.~\ref{fig: XX} for various defect strengths $\lambda$.
The results are compared to the fermionic negativity $\lnf$, shown by the empty symbols,
and indicate that the upper bound in \eqref{ub} actually holds even without the additional
constant, i.e. one has $\lns \le \lnf$. The two quantities have a very similar qualitative
behaviour, with their difference diminishing with decreasing $\lambda$.
Unfortunately, however, the simulation times as well as the size of the segments
are severely limited in the tDMRG simulations due to the increasing entanglement and
bond dimension during time evolution, especially for higher values of $\lambda$.
This makes a quantitative analysis of the discrepancy between $\lns$ and $\lnf$
rather complicated.

%%%%%%%%%%%%%%%%%%%%%%%%%%%%%%%%%%%%%%%%%%%
%%%%%%%%%%%%%%%%%%%%%%%%%%%%%%%%%%%%%%%%%%%

\begin{figure}[t]
  \centering
    \hspace*{-0.25cm}
%   \subfloat{\includegraphics{./PLOTS/PDF/XXZ.pdf}}
%   \subfloat{\includegraphics{./PLOTS/PDF/XXZ_defect_del05.pdf}} 
    \includegraphics[width=0.49\columnwidth]{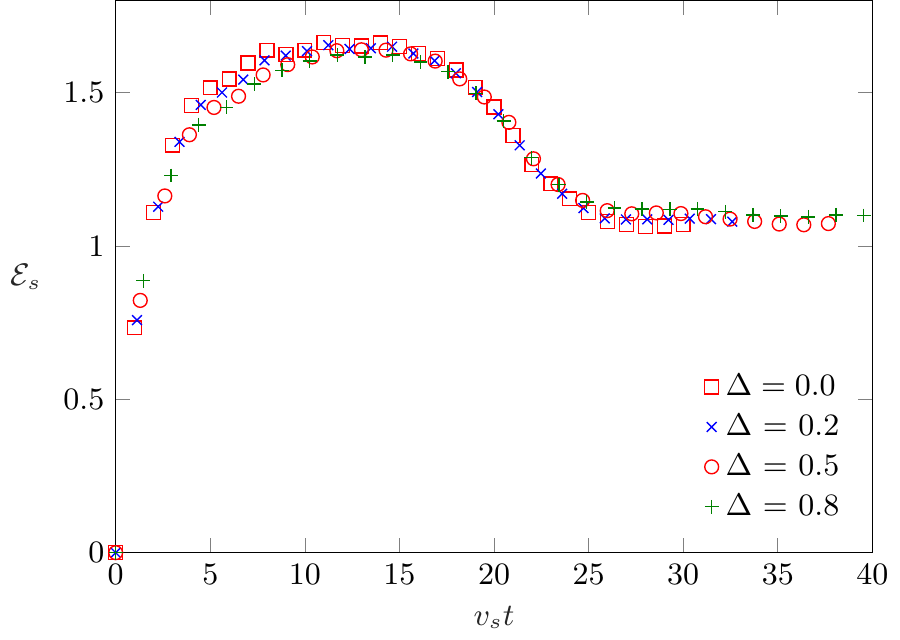}
    \includegraphics[width=0.49\columnwidth]{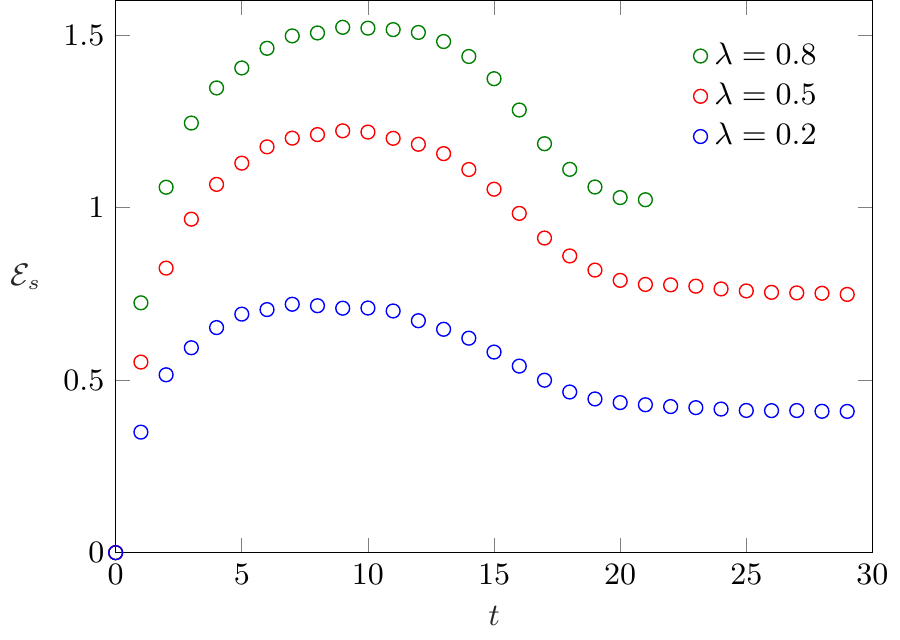}
  \caption{Time evolution of the entanglement negativity $\lns$ for two adjacent segments
  of size $\ell = 20$ in an XXZ chain with $2N=100$.
  Left: homogeneous quench ($\lambda=1$) for different anisotropies $\Delta$.
  Right: quench across a defect of different strengths $\lambda$ and $\Delta = 0.5$.}
  \label{fig: XXZ}
\end{figure}

We also performed analogous tDMRG simulations for the XXZ chain with the
anisotropy parameter $0< \Delta < 1$. The fermionic analogue of this setting
corresponds to an interacting problem and thus not amenable to Gaussian techniques.
We first considered the homogeneous case $\lambda=1$,
where the post-quench Hamiltonian is integrable and its low-energy behaviour
is described by a Luttinger liquid. In particular, the spreading of excitations
created above the ground state is given by the spinon velocity \cite{Franchini}
\eq{
v_s = \frac{\pi}{2}\frac{\sqrt{1-\Delta^2}}{\acos (\Delta)} \, .
}
This strongly suggests that the main difference with respect to the homogeneous
XX quench is due to the change in the Fermi velocity.
On the left of Fig.~\ref{fig: XXZ} we have thus plotted the logarithmic negativity $\lns$
calculated for various $\Delta$ against the variable $v_s t$, which indeed leads to a
nice data collapse.

The situation for $\lambda\ne1$ is more complicated, as the presence of the defect
breaks the integrability of the model. The time evolution of the negativity
is shown on the right of Fig.~\ref{fig: XXZ} for various defect strengths $\lambda$
and fixed $\Delta=0.5$, for a segment size $\ell=20$. Qualitatively, one observes
a very similar behaviour as for the XX chain in Fig.~\ref{fig: XX}. However, in previous
studies of the half-chain entropy in Ref.~\cite{collura2013entanglement} it was observed
that the entropy growth is actually suppressed for repulsive interactions $\Delta > 0$,
corresponding to an effective central charge that goes to zero in the limit of large chain
sizes. This is actually the same mechanism that was found for the ground-state entropy
of the XXZ chain with a defect \cite{zhao2006critical}, and is the manifestation of a
Kane-Fisher type renormalization behaviour \cite{KF92}.
Therefore it is reasonable to expect that the entanglement negativity would
show a similar behaviour in the limit of large $\ell$. Unfortunately, however,
the segment sizes required to test such a crossover are well beyond the limitations
of our simulations.

%%%%%%%%%%%%%%%%%%%%%%%%%%%%%%%%%%%%%%%%%%%
%%%%%%%%%%%%%%%%%%%%%%%%%%%%%%%%%%%%%%%%%%%

\section{Discussion\label{sec:disc}}

We have studied entanglement in a hopping chain with a defect,
focusing on the fermionic version of the logarithmic negativity $\lnf$ between two segments
neighbouring the defect, and its relation to the R\'enyi mutual information $\mi$.
In the ground state of the chain, the negativity scales logarithmically with an
effective central charge $\ceff$, and the difference $\lnf - \mi$ goes to zero for increasing
segment sizes. For a quench across the defect, starting from disconnected half-chains both
at half filling, the growth of the negativity is logarithmic in time and the prefactor
seems to be well approximated by $\ceff$. When the quench is performed from biased fillings,
the entanglement growth becomes linear, followed by a saturation at an extensive value,
which is due to backscattering from the defect and can be understood in a semiclassical picture.
Although the ansatz \eqref{lndwsc} gives a very good leading order description
of both $\lnf$ and $\mi$, the subleading corrections behave differently and
their difference in the steady state remains finite even for large segment sizes.
We have also calculated the standard logarithmic negativity $\lns$ via DMRG methods
in the XXZ spin chain after a (unbiased) quench across a defect.
In the noninteracting XX case, closely related to the fermionic chain,
we found that the spin-chain negativity is upper bounded by the fermionic one, $\lns\le\lnf$.
In the general XXZ case, the results for $\lns$ look qualitatively similar to the XX case
for the small segment sizes attainable.

While the entanglement growth in the biased case has a very clear physical interpretation,
the result for equal fillings is harder to grasp and would require some insight from CFT calculations.
In fact, for the bipartite case of a half-chain, the CFT representation of the density matrix after the
quench can be transformed into the one for the ground state by an appropriate conformal mapping
\cite{WWR18}. Unfortunately, however, this transformation works only for the half-chain
and it is unclear whether a generalization to our geometry exists.

It is important to stress that, although the semiclassical picture for unequal fillings
is analogous to the one for a global quench \cite{alba2019quantuminfo}, the qualitative
behaviour of the negativity is completely different. Indeed, in the latter case the
quasiparticles are created only at $t=0$, and thus the pairs that contribute to the
entanglement growth eventually leave the segments. This implies that the negativity
will decrease again for large times, decaying towards zero. In contrast, in our case
there is a constant production of entangled pairs at the defect, and thus the negativity
keeps growing until it eventually saturates at an extensive value. Furthermore, while
for the global quench the deviation between $\lnf$ and $\mi$ seems to vanish for
increasing $\ell$, for the defect we observe a finite difference between the two quantities.
Understanding the origin of this discrepancy requires further investigations.

Finally, it would be interesting to extend these investigations to disjoint segments.
In particular, it would be illuminating to see how the disagreement between the
fermionic and XX chain negativities changes with separation.
One expects the discrepancy to become larger, as the partial transpose is a sum
of four fermionic Gaussian operators already in the ground state \cite{CTC15}.
While the extension of both the fermionic as well as the spin-chain calculations are,
in principle, straightforward, the computational effort of the DMRG calculations
are much more demanding and are thus left for future studies.

\begin{acknowledgments}

We thank D. Bauernfeind, F. Maislinger and Z. Zimbor\'as for useful discussions.
The authors acknowledge funding from the Austrian Science Fund (FWF) through
project No. P30616-N36.

\end{acknowledgments}

\newpage

\onecolumngrid

\appendix
\section{Correlation matrices for the defect\label{app:corr}}

We collect here the integral formulas for the correlation matrix elements
$\braket{c^\dag_m c_n}$ in the thermodynamic limit $N\to\infty$ of the hopping chain \eqref{eq: H},
with a single weak hopping defect parametrized by $\lambda=\ee^{-\nu}$.
We consider both the ground state of the chain as well as the NESS
after time evolution from a domain wall. The former has been considered
in Ref. \cite{peschel2005entanglementdefect} while in the latter case the results were obtained in
Ref. \cite{LSP19}. In each case the result depends on whether the
lattice sites are chosen on the same (left or right) or opposite sides of the defect,
i.e. the correlation matrix has a block form. We shall only consider matrix
elements with $m \le n$, since the others follow from hermiticity.

In the ground state of an infinite chain, the matrix elements read
\eq{
C_{m,n} =
\begin{cases}
C_0(n-m) - C_1(n+m) & \mbox{if $m,n\ge 1$} \\
C_0(n-m) - C_1(2-n-m) & \mbox{if $m,n\le 0$} \\
C_2(n-m) & \mbox{if $m\le 0$, $n \ge 1$} 
\end{cases}
\label{cgs}}
where the different contributions depend only on the
difference $r=n-m$ and sum $s=n+m$ of the indices.
The first translationally invariant piece is given by
\eq{
C_0(r) = \frac{\sin (\frac{\pi}{2}r)}{\pi r} \, ,
}
which is just the homogeneous result. The extra contributions on the same side
of the defect were obtained in \cite{peschel2005entanglementdefect} and read
\eq{
C_1(s) = \frac{\sh \, \nu}{2} (\ee^\nu I_{s} - \ee^{-\nu} I_{s-2}),
}
where
\eq{
I_s = \int_{0}^{\pi/2} \frac{\dd q}{\pi}
\frac{\cos q s}{\sh^2 \, \nu + \sin^2 q} \, .
}
Interestingly, one observes that the contributions $C_1(2k)=0$ vanish completely
for even $s=2k$, which follows from the property $\ee^\nu I_{2k} = \ee^{-\nu} I_{2k-2}$
of the integrals defined above.
Finally, the matrix elements on opposite sides of the defect
are given by
\eq{
C_2(r)=\frac{\ch \, \nu}{2} I_r 
- \frac{\ee^{\nu}}{4} I_{r+2} - \frac{\ee^{-\nu}}{4} I_{r-2} \, .
}
Using the property found above, it is easy to see that also these
contributions vanish for $r=2k$. Thus the correlation matrix has a
checkerboard structure as in the homogeneous case.

In the limit $\nu=0$ one has trivially $C_1(s) = 0$, whereas for the
offdiagonal block one finds
\eq{
C_2(r) = \int_{0}^{\pi/2} \frac{\dd q}{2\pi}
\frac{\cos q r - \left[\cos q(r-2)+\cos q(r+2)\right]/2}{\sin^2 q}=
\int_{0}^{\pi/2} \frac{\dd q}{\pi}
\cos q r= C_0(r) \, ,
}
i.e. one recovers the results for the homogeneous chain.
In the opposite limit $\nu \to \infty$ of a vanishing defect coupling,
one finds $C_1(s)=C_0(s)$ which is the result for a half-infinite chain.
On the other hand, $C_2(r)=0$ as it should for two decoupled half-chains.

Now we consider the correlation matrix elements $\tilde C_{m,n}$
for the NESS, which emerges in the $t \to \infty$ limit of time
evolution from a domain wall initial state.
The results were obtained in \cite{LSP19} by solving the problem for a
finite system size and time and then considering
the limits $N \to \infty$ and $t \to \infty$ via contour integration tricks.
In order to bring the formulas for the matrix elements
in a transparent form, it is useful to introduce the transmission and
reflection coefficients
\eq{
T_q = \frac{\sin^2 q}{\sh^2 \, \nu + \sin^2 q} \, , \qquad
R_q = \frac{\sh^2 \, \nu}{\sh^2 \, \nu + \sin^2 q} \, ,
}
as well as the auxiliary integral expressions
\eq{
\tilde I_r = \int_{0}^{\pi} \frac{\dd q}{2\pi}
\frac{i \, \sin q r}{\sh^2 \, \nu + \sin^2 q} \, .
}
%
%Note that $\tilde I_s$ differs from $I_s$ only in the limits of integration and a factor 2.
%Considering the symmetry properties of the integrands under the reflection $q \to \pi-q$,  one has
%$\tilde I_{2k-1}=0$ as well as $\tilde I'_{2k}=0$ for any integer $k$.
 
Analogously to \eqref{cgs}, the matrix elements can be written down by a 
separation of cases
\eq{
\tilde C_{m,n} =
\begin{cases}
\tilde C_0(n-m) & \mbox{if $m,n\ge 1$} \\ %+ \tilde C_1(n+m)
\delta_{m,n} - \tilde C_0(m-n) & \mbox{if $m,n\le 0$} \\ %- \tilde C_1(2-m-n)
\tilde C_2(n-m) + \tilde C_3(n+m) & \mbox{if $m\le 0$, $n \ge 1$} 
\end{cases}
\label{cness}}
The correlations are thus translationally invariant if both sites
are located on the same side. They can be written in a very instructive form
\eq{
\tilde C_0(r) = 
\int_{0}^{\pi} \frac{\dd q}{2\pi} T_q \, \ee^{i q r}.
\label{c0ness}}
Indeed, this can be interpreted as a correlation matrix where the
occupation function is given by the transmission probability for all the
modes with positive velocities.
Note that the correlations on the left/right hand side are now related
by the symmetry property $\tilde C_{m,n}=\delta_{m,n} - \tilde C_{1-m,1-n}$.
If the sites are located on opposite sides, the correlations are given via
the expressions
\eq{
\tilde C_2(r) =% i \int_{0}^{\pi} \frac{\dd q}{2\pi}
%\left[\ch \, \nu \, T_q \sin qr - \sqrt{T_qR_q} \cos q \cos qr \right] \, ,
\frac{\ch \, \nu}{2} \tilde I_r 
- \frac{\ee^{\nu}}{4} \tilde I_{r+2} - \frac{\ee^{-\nu}}{4} \tilde I_{r-2} \,
}
and 
\eq{
\tilde C_3(s) = i \int_{0}^{\pi} \frac{\dd q}{2\pi} \sqrt{T_qR_q} \, \ee^{iq(s-1)} \, .
}

The limiting cases are also straightforward to obtain.
For $\nu=0$ we have $T_q \equiv 1$ and $R_q \equiv 0$
such that
\eq{
\tilde C_0(r) =
\frac{1}{2} \delta_{r,0} + i \int_{0}^{\pi} \frac{\dd q}{2\pi} \sin qr \, ,
}
which is exactly the NESS result for the homogeneous chain. For the offdiagonal
block one obtains $C_3(s)=0$ as well as
\eq{
\tilde C_2(r) = i\int_{0}^{\pi} \frac{\dd q}{4\pi}
\frac{\sin q r - \left[\sin q(r-2)+\sin q(r+2)\right]/2}{\sin^2 q}=
i\int_{0}^{\pi} \frac{\dd q}{2\pi}
\sin q r \, ,
}
i.e. the full matrix becomes translationally invariant, as it should.
In the opposite limit $\nu\to\infty$ one has $T_q \equiv 0$ and $R_q \equiv 1$,
and thus $\tilde C_0(r)=\tilde C_2(r)=\tilde C_3(s)=0$. One then simply
recovers the initial $t=0$ form of the correlation matrix since the
transmission vanishes between the two half-chains.

\section{CFT treatment of the domain wall quench\label{app:cft}}

In this appendix we present the CFT calculation of the mutual information
and entanglement negativity for a domain wall initial state time evolved
with a homogeneous hopping chain. The key insight to the problem
was provided in Ref. \cite{DSVC17}, where it was shown that the inhomogeneous
time-evolved state can be mapped onto a CFT with a curved background metric.
This metric was first obtained in \cite{ADSV16} where the imaginary-time evolution
of the domain wall initial state was considered. Alternatively, one could
use the exact mapping from the domain-wall melting to a ground-state problem
with a linear potential \cite{EIP09,VIR17}. Here we will follow the latter route.

Let us consider a free-fermion chain with a slowly varying linear potential,
with the corresponding length scale given by $t$. The ground state of this
chain is unitarily equivalent to the time-evolved state starting from a domain wall \cite{EIP09}.
For $t \gg 1$ one can apply a local density approximation (LDA), i.e. one assumes
that the ground state around position $x$ is locally equivalent to a
homogeneous ground state, corresponding to the dispersion $\omega_q = -\cos q + x/t$.
The spatially varying Fermi momentum and velocity are then given by, respectively,
\eq{
q_F(x)=\arccos(x/t) \, ,
\qquad
v_F(x)=\sin q_F(x)=\sqrt{1-(x/t)^2} \, .
}

The LDA yields a description where the state can be \emph{locally} described
via a 2D massless Dirac fermion field theory. The crucial finding of Ref. \cite{DSVC17} is that
one can define a \emph{globally} valid Dirac theory, living in a curved background
metric, where the changing of the Fermi velocity can be absorbed by introducing
the coordinate transformation
\eq{
z = \int_{0}^{x} \frac{\dd x'}{v_F(x')} + i y = t \arcsin \frac{x}{t} + i y \, .
\label{z}}
The curved metric is then given by $\dd s^2 = \ee^{2\sigma(x)} \dd z \dd \bar z$,
where the Weyl factor has to be chosen as $\ee^{\sigma(x)} = v_F(x)$ in order
to reproduce the local fermion propagators.

Once the proper metric and field theory have been identified, the calculation of
the entropy can be performed by applying the replica trick and the corresponding
twist-field formalism \cite{CC09}. Namely, the R\'enyi entropy $S_n$ can be obtained
by calculating expectation values of twist fields $\mathcal{T}_n$ and $\mathcal{\bar T}_n$
inserted at the spatial boundaries (and imaginary time $y=0$) of the subsystem at hand.
Here we focus on an interval $A_1=\left[x_1,x_2\right]$ such that
%
%\eq{
%S_n  = \frac{1}{1-n} \ln \Tr (\rho^n_A)
%}
%
\eq{
\Tr (\rho^n_{A_1})=\epsilon(x_1)^{\Delta_n} \, \epsilon(x_2)^{\Delta_n}
\braket{\mathcal{T}_n(x_1) \mathcal{\bar T}_n(x_2)} ,
%S_n  = \frac{1}{1-n} \ln \, 
\label{trtw}
}
where the scaling dimension of the twist fields is given by
\eq{
\Delta_n = \frac{c}{12}\left(n-\frac{1}{n}\right) ,
\label{scd}}
with the central charge being $c=1$ for the Dirac theory.
Note that we have explicitly included a UV cutoff $\epsilon(x)$ in \eqref{trtw} which,
in contrast to homogeneous systems, carries a spatial dependence and thus
cannot be ignored. Indeed, since the only relevant microscopic energy scale on
the lattice is given by the Fermi velocity, the cutoff must be chosen as
$\epsilon(x)=\epsilon_0 \, v^{-1}_F(x)$, where $\epsilon_0$ is a dimensionless constant.

In order to evaluate the expectation value in \eqref{trtw}, one should point out that,
due the change of coordinates in \eqref{z}, the curved-space field theory lives on
the infinite strip $\left[-\frac{\pi}{2}t,\frac{\pi}{2}t\right] \times \mathbb{R}$.
Therefore, one has to first map the theory onto the upper half plane by the
conformal transformation $g(z)=\ee^{i(z/t+\pi/2)}$. The twist-field two-point function
can then be written as
\eq{
\braket{\mathcal{T}_n(x_1) \mathcal{\bar T}_n(x_2)}=
%\prod_{i=1}^2
%\left(\epsilon_1 \, \ee^{-\sigma(z_1)}\left|\frac{\dd g(z_1)}{\dd z_1}\right|\right)^{\Delta_n}
\left(\ee^{-\sigma(x_1)}\left|\frac{\dd g(z_1)}{\dd z_1}\right|\right)^{\Delta_n}
\left(\ee^{-\sigma(x_2)}\left|\frac{\dd g(z_2)}{\dd z_2}\right|\right)^{\Delta_n}
\braket{\mathcal{T}_n(g(z_1)) \mathcal{\bar T}_n(g(z_2))}_{\mathrm{UHP}} \, .
}
In the above expression we simply used the transformation properties of
the twist fields under the Weyl transformation (i.e. changing to the curved-space coordinates)
as well as the mapping $g(z)$. The remaining step is to evaluate the two-point
function on the upper half plane which, using the method of images, can be
written as a four-point function on the full plane. Up to multiplicative constants,
one obtains for the Dirac theory \cite{DSVC17}
%
%\eq{
%\braket{\mathcal{T}_n(z)}=
%\left(\epsilon \, \ee^{-\sigma(z)}\left|\frac{\dd g(z)}{\dd z}\right|\right)^{\Delta_n}
%\braket{\mathcal{T}_n(g(z))}_{\mathrm{UHP}} \, ,
%\qquad
%\braket{\mathcal{T}_n(g(z))}_{\mathrm{UHP}}=
%\left[ \mathrm{Im} \, g(z) \right]^{-\Delta_n}
%}
%
\eq{
\braket{\mathcal{T}_n(g(z_1)) \mathcal{\bar T}_n(g(z_2))}_{\mathrm{UHP}}=
\left[
%\frac{\mathrm{Im} \, g(z_1) \, \mathrm{Im} \, g(z_2)}{\epsilon_1 \, \epsilon_2}
\mathrm{Im} \, g(z_1) \, \mathrm{Im} \, g(z_2) \, \eta_{1,2} \right]^{-\Delta_n} \, ,
\qquad
\eta_{1,2}=\frac{|g(z_1)-g(z_2)|^2}{|g(z_1)-g^*(z_2)|^2} \, .
}
It should be noted that, for a generic CFT, the result is more complicated
and is multiplied by a non-universal function $\mathcal{F}(\eta_{1,2})$
of the four-point ratio $\eta_{1,2}$, see e.g. \cite{CCT09}. For the Dirac theory,
however, one has $\mathcal{F} \equiv 1$ \cite{CFH05}.

We are now ready to calculate the R\'enyi mutual information \eqref{salpha} between
two adjacent intervals $A_1=\left[x_1,x_2\right]$ and $A_2=\left[x_2,x_3\right]$.
Putting everything together, one arrives at the result
\eq{
\mathcal{I}_n=
\frac{1+n}{6n} \ln
\left[\epsilon^{-1}(x_2)\,\ee^{\sigma(x_2)}\left|\frac{\dd g(z_2)}{\dd z_2}\right|^{-1}
\mathrm{Im} \, g(z_2)
\left(\frac{\eta_{1,2} \, \eta_{2,3}}{\eta_{1,3}}\right)^{1/2}\right].
\label{micft}}

The calculation for the logarithmic negativity follows a similar procedure,
but is slightly more involved. In the replica approach it can be written as
\cite{CCT12,CCT13}
\eq{
\lneg = \lim_{n_e \to 1} \ln \Tr (\rho_A^{T_2})^{n_e} \, ,
\label{lnrep}}
where the calculation has to be carried out for an even $n_e$ number of
replicas and then taking the limit $n_e\to1$. Indeed, the limit $n_o \to 1$
from an odd number of copies would give the log of the trace (which is trivially zero)
instead of the trace norm. Furthermore, the effect of the
partial transpose is to interchange the twist operators $\mathcal{T}_n$ and $\mathcal{\bar T}_n$
located at the ends of the segment $A_2$ over which the transpose is taken.
We will restrict ourselves to adjacent intervals $A_1=\left[x_1,x_2\right]$
and $A_2=\left[x_2,x_3\right]$, such that the trace can be written as the three-point function
\eq{
\Tr (\rho_A^{T_2})^{n} =
%\epsilon(z_1)^{\Delta_{(1)}} \, \epsilon(z_2)^{\Delta_{(2)}} \, \epsilon(z_3)^{\Delta_{(3)}}
\prod_{i=1}^3\epsilon(x_i)^{\Delta_{(i)}} 
\braket{\mathcal{T}_n(x_1) \mathcal{\bar T}^2_n(x_2) \mathcal{T}_n(x_3)} .
}
where the scaling dimensions are given by
\eq{
\Delta_{(1)} =\Delta_{(3)}=\Delta_n \, , \qquad
\Delta_{(2)} =
\begin{cases}
\Delta_{n_o} & n=n_o \, , \\
2\Delta_{n_e/2} & n=n_e \, .
\end{cases}
\label{scd2}}
Clearly, the scaling dimension $\Delta_{(2)}$ corresponding to the composite field
$\mathcal{\bar T}^2_n$ shows a strong parity dependence.

To calculate the twist-field expectation value, one uses again the transformation properties 
\eq{
\braket{\mathcal{T}_n(x_1) \mathcal{\bar T}^2_n(x_2) \mathcal{T}_n(x_3) }=
\prod_{i=1}^3 \left(\ee^{-\sigma(x_i)}\left|\frac{\dd g(z_i)}{\dd z_i}\right|\right)^{\Delta_{(i)}}
\braket{\mathcal{T}_n(g(z_1)) \mathcal{\bar T}^2_n(g(z_2)) \mathcal{T}_n(g(z_3))}_{\mathrm{UHP}} \, .
}
The last step is to evaluate the three-point function on the upper half plane,
which has already been considered in \cite{wen2015entanglement}.
For the Dirac theory one has
\eq{
\braket{\mathcal{T}_n(g(z_1)) \mathcal{\bar T}^2_n(g(z_2)) \mathcal{T}_n(g(z_3))}_{\mathrm{UHP}}=
%\frac{\mathrm{Im} \, g(z_1) \, \mathrm{Im} \, g(z_2)}{\epsilon_1 \, \epsilon_2}
\prod_{i=1}^3 \left[ \mathrm{Im} \, g(z_i) \right]^{-\Delta_{(i)}} 
\left[ \eta_{1,2}^{\Delta_{(2)}} \eta_{2,3}^{\Delta_{(2)}}
\eta_{1,3}^{\Delta_{(1)}+\Delta_{(3)}-\Delta_{(2)}} \right]^{-1/2},
}
where the four-point ratios are defined as
\eq{
\eta_{i,j}=\frac{|g(z_i)-g(z_j)|^2}{|g(z_i)-g^*(z_j)|^2} \, .
}
Note that here we assumed that the non-universal function
$\mathcal{F}(\left\{\eta_{i.j}\right\})$, which could depend on the full
operator content for a generic CFT, becomes again trivial for the Dirac theory \cite{CFH05}.
Finally, since $\Delta_{1}=0$ from \eqref{scd}, the only nontrivial scaling
dimension from \eqref{scd2} that survives the replica limit \eqref{lnrep} is
$\lim_{n_e \to 1} \Delta_{(2)}=2\Delta_{1/2}=-1/4$. In turn, the entanglement
negativity can be written as
\eq{
\lneg = \frac{1}{4} \ln
\left[\epsilon^{-1}(x_2)\,\ee^{\sigma(x_2)}\left|\frac{\dd g(z_2)}{\dd z_2}\right|^{-1}
\mathrm{Im} \, g(z_2)
\left(\frac{\eta_{1,2} \, \eta_{2,3}}{\eta_{1,3}}\right)^{1/2}\right].
\label{lncft}}

Comparing \eqref{lncft} to \eqref{micft}, one finds immediately $\lneg=\mi$.
Note, however, that the result of the CFT calculation is valid only up to a
nonuniversal additive constant. Nevertheless, one can use a continuity argument
to make sure that this constant is the same for both quantities. Namely,
if one considers a \emph{bipartite} situation where $A$ is the full system,
then one has exactly $\lneg=\mi$. Therefore the equality should be valid,
up to subleading terms, for arbitrary adjacent segments.

Finally, it should be stressed that the calculation was carried out in complete
generality for free-fermion systems that have an underlying curved-space CFT.
In the last step we apply the result to our specific example of the linear potential.
Introducing the scaling variables $\xi_i=x_i/t$, the various factors appearing in the
argument of \eqref{lncft} read
\eq{
\epsilon^{-1}(x_2) = \ee^{\sigma(x_2)} = \mathrm{Im} \, g(z_2) =
\sqrt{1-\xi_2^2} \, , \qquad
\left|\frac{\dd g(z_2)}{\dd z_2}\right|^{-1} = t \, ,
}
whereas the square-root of the four-point ratios can be evaluated as
\eq{
(\eta_{i,j})^{1/2} = \frac{1-\xi_i\xi_j-\sqrt{(1-\xi_i^2)(1-\xi_j^2)}}{|\xi_i-\xi_j|} \, .
}
In particular, for the symmetric arrangement of the segments considered
in the main text, $\xi_3=-\xi_1=\xi=\ell/t$ and $\xi_2=0$, the ratios further
simplify to
\eq{
(\eta_{1,2})^{1/2}=(\eta_{2,3})^{1/2}=\frac{1-\sqrt{1-\xi^2}}{\xi} \, ,
\qquad
(\eta_{1,3})^{1/2}=\xi \, ,
}
and plugging into \eqref{lncft} yields the result in \eqref{lndwhom}.

\bibliography{literatur}

\end{document}